\documentclass[reprint, nofootinbib, prb, superscriptaddress]{revtex4-1}
\usepackage{amsmath} \usepackage{graphicx} \usepackage{dcolumn}
\usepackage{bm} 
\usepackage{amssymb}
\usepackage{pstricks}
\usepackage{xr}

\begin{document}

\newlength{\figurewidth}
\setlength{\figurewidth}{\columnwidth}

\newcommand{\prtl}{\partial}
\newcommand{\la}{\left\langle}
\newcommand{\ra}{\right\rangle}
\newcommand{\dla}{\la \! \! \! \la}
\newcommand{\dra}{\ra \! \! \! \ra}
\newcommand{\we}{\widetilde}
\newcommand{\smfp}{{\mbox{\scriptsize mfp}}}
\newcommand{\smp}{{\mbox{\scriptsize mp}}}
\newcommand{\sph}{{\mbox{\scriptsize ph}}}
\newcommand{\sinhom}{{\mbox{\scriptsize inhom}}}
\newcommand{\sneigh}{{\mbox{\scriptsize neigh}}}
\newcommand{\srlxn}{{\mbox{\scriptsize rlxn}}}
\newcommand{\svibr}{{\mbox{\scriptsize vibr}}}
\newcommand{\smicro}{{\mbox{\scriptsize micro}}}
\newcommand{\scoll}{{\mbox{\scriptsize coll}}}
\newcommand{\sattr}{{\mbox{\scriptsize attr}}}
\newcommand{\sth}{{\mbox{\scriptsize th}}}
\newcommand{\sauto}{{\mbox{\scriptsize auto}}}
\newcommand{\seq}{{\mbox{\scriptsize eq}}}
\newcommand{\teq}{{\mbox{\tiny eq}}}
\newcommand{\sinn}{{\mbox{\scriptsize in}}}
\newcommand{\suni}{{\mbox{\scriptsize uni}}}
\newcommand{\tin}{{\mbox{\tiny in}}}
\newcommand{\scr}{{\mbox{\scriptsize cr}}}
\newcommand{\tstring}{{\mbox{\tiny string}}}
\newcommand{\sperc}{{\mbox{\scriptsize perc}}}
\newcommand{\tperc}{{\mbox{\tiny perc}}}
\newcommand{\sstring}{{\mbox{\scriptsize string}}}
\newcommand{\stheor}{{\mbox{\scriptsize theor}}}
\newcommand{\sGS}{{\mbox{\scriptsize GS}}}
\newcommand{\sBP}{{\mbox{\scriptsize BP}}}
\newcommand{\sNMT}{{\mbox{\scriptsize NMT}}}
\newcommand{\sbulk}{{\mbox{\scriptsize bulk}}}
\newcommand{\tbulk}{{\mbox{\tiny bulk}}}
\newcommand{\sXtal}{{\mbox{\scriptsize Xtal}}}
\newcommand{\sliq}{{\text{\tiny liq}}}

\newcommand{\smin}{\text{min}}
\newcommand{\smax}{\text{max}}

\newcommand{\saX}{\text{\tiny aX}}
\newcommand{\slaX}{\text{l,{\tiny aX}}}

\newcommand{\svap}{{\mbox{\scriptsize vap}}}
\newcommand{\sjam}{J}
\newcommand{\Tm}{T_m}
\newcommand{\sTS}{{\mbox{\scriptsize TS}}}
\newcommand{\sDW}{{\mbox{\tiny DW}}}
\newcommand{\cN}{{\cal N}}
\newcommand{\cB}{{\cal B}}
\newcommand{\br}{\bm r}
\newcommand{\be}{\bm e}
\newcommand{\cH}{{\cal H}}
\newcommand{\cHlt}{\cH_{\mbox{\scriptsize lat}}}
\newcommand{\sthermo}{{\mbox{\scriptsize thermo}}}

\newcommand{\bu}{\bm u}
\newcommand{\bk}{\bm k}
\newcommand{\bX}{\bm X}
\newcommand{\bY}{\bm Y}
\newcommand{\bA}{\bm A}
\newcommand{\bb}{\bm b}

\newcommand{\lintf}{l_\text{intf}}

\newcommand{\DV}{\delta V_{12}}
\newcommand{\sout}{{\mbox{\scriptsize out}}}
\newcommand{\dv}{\Delta v_{1 \infty}}
\newcommand{\dvin}{\Delta v_{2 \infty}}

\newcommand{\wtp}{\tilde{p}}
\newcommand{\wtK}{\widetilde{K}}
\newcommand{\wtgm}{\tilde{\gamma}}
\newcommand{\wtg}{\widetilde{g}}

\def\Xint#1{\mathchoice
   {\XXint\displaystyle\textstyle{#1}}%
   {\XXint\textstyle\scriptstyle{#1}}%
   {\XXint\scriptstyle\scriptscriptstyle{#1}}%
   {\XXint\scriptscriptstyle\scriptscriptstyle{#1}}%
   \!\int}
\def\XXint#1#2#3{{\setbox0=\hbox{$#1{#2#3}{\int}$}
     \vcenter{\hbox{$#2#3$}}\kern-.5\wd0}}
\def\ddashint{\Xint=}
\def\dashint{\Xint-}
\title{Aging, jamming, and the limits of stability of amorphous
  solids}

\author{Vassiliy Lubchenko} \email{vas@uh.edu}
\affiliation{Departments of Chemistry and Physics, University of
  Houston, Houston, TX 77204}

\author{Peter G. Wolynes} \affiliation{Departments of Chemistry,
  Physics and Astronomy, and Center for Theoretical Biological
  Physics, Rice University, Houston, TX 77005}

\date{\today}

\begin{abstract}

  Apart from not having crystallized, supercooled liquids can be
  considered as being properly equilibrated and thus can be described
  by a few thermodynamic control variables.  In contrast, glasses and
  other amorphous solids can be arbitrarily far away from equilibrium
  and require a description of the history of the conditions under
  which they formed. In this paper we describe how the locality of
  interactions intrinsic to finite-dimensional systems affects the
  stability of amorphous solids far off equilibrium. Our analysis
  encompasses both structural glasses formed by cooling and colloidal
  assemblies formed by compression. A diagram outlining regions of
  marginal stability can be adduced which bears some resemblance to
  the quasi-equilibrium replica meanfield theory phase diagram of hard
  sphere glasses in high dimensions but is distinct from that
  construct in that the diagram describes not true phase transitions
  but kinetic transitions that depend on the preparation protocol.
  The diagram exhibits two distinct sectors. One sector corresponds to
  amorphous states with relatively open structures, the other to high
  density, more closely-packed ones. The former transform rapidly
  owing to there being motions with no free energy barriers; these
  motions are string-like locally.  In the dense region, amorphous
  systems age via compact activated reconfigurations.  The two regimes
  correspond, in equilibrium, to the collisional or uniform liquid and
  the so called landscape regime, respectively. These are separated by
  a spinodal line of dynamical crossovers. Owing to the rigidity of
  the surrounding matrix in the landscape, high-density part of the
  diagram, a sufficiently rapid pressure quench adds compressive
  energy which also leads to an instability toward string-like motions
  with near vanishing barriers. Conversely, a dilute collection of
  rigid particles, such as a colloidal suspension leads, when
  compressed, to a spatially heterogeneous structure with percolated
  mechanically-stable regions. This jamming corresponds to the onset
  of activation when the spinodal line is traversed from the low
  density side. We argue that a stable glass made of sufficiently
  rigid particles can also be viewed as exhibiting sporadic and
  localized buckling instabilities that result in local jammed
  structures. The lines of instability we discuss resemble the Gardner
  transition of meanfield systems but, in contrast, do not result in
  true criticality owing to being short-circuited by activated
  events. The locally marginally stable modes of motion in amorphous
  solids correspond to secondary relaxation processes in structural
  glasses. Their relevance to the low temperature anomalies in glasses
  is also discussed.

\end{abstract}

\maketitle


{\hfill \em Fluctuat nec mergitur.}

{\hfill Motto of Paris, France}

\section{Introduction}

Rigidity can emerge, seemingly miraculously, from the interaction of
many pieces that are loosely connected: Tents, arches, and domes are
among the earliest inventions of humans. Likewise, small fluctuations
can sometimes cause such structures to collapse without
warning. Although statics has been part of the architectural
curriculum since ancient times, only in the 19th century did Maxwell
write down general criteria for building a stable structure out of
many unmoving but loosely connected parts.~\cite{MaxwellConstraint}
His ideas have resurfaced in the last decades in the description of
amorphous solids such as granular assemblies, at the macroscopic
level, and glasses at the molecular size scale.~\cite{PHILLIPS1985699}
This modern complex of ideas emphasizes the importance of a ``jammed
state'' of matter,~\cite{LiuNagelSaarlosWyart} which describes an
assembly that just barely holds together (i.e. is ``marginally
stable'') but does so under the constraint of a large external
pressure.

An apparently distinct view of the emergence of rigidity in amorphous
solids more suited to the molecular scale has also emerged, based on
the notions of aperiodic crystals and the complexity of the free
energy landscapes of disordered systems that have a thermodynamically
large number of aperiodic minima. This view comes to terms with the
fact that the molecular constituents of amorphous solids are in a
state of cease-less atomic motion. The availability of many possible
states of repose for static heaps of the constituent particles then
not only raises the possibility of flow between these states but also
makes such flow inevitable at any finite temperature. Thus Maxwell's
criteria for engineering macroscopic structures need to be
supplemented, in such molecular systems, by considerations of dynamics
and thermodynamics. We see that the rigidity of glasses must be only
apparent: Flow simply becomes so slow that an amorphous solid acts as
a rigid body on human timescales and the puzzle of rigidity becomes
the puzzle of explaining the origin of those extraordinary
timescales. The most comprehensive quantitative explanation of these
long timescales of motion in amorphous solids is provided by the
Random First Order Transition (RFOT) theory of glasses.~\cite{LW_ARPC,
  L_AP} This theory explains that the slowness arises from the
difficulty of finding ways of rearranging loose regions from one
jumbled state of repose (``an aperiodic crystal'') to another such
mechanically stable configuration. Intermittently transiting from one
aperiodic crystal arrangement to another, the system can be called a
``glassy liquid.'' This difficulty of finding alternative aperiodic
structures grows as the thermal motion diminishes and thereby
diminishes the availability of new configurations to which one can
move, an impending entropy crisis.~\cite{Kauzmann} If one could reach
this entropy crisis, a true phase transition into a unique but
aperiodic crystal is envisioned to occur.

The RFOT theory relies on the locality of interactions in order to
understand the finite time dynamics but a great deal of progress has
been made formalizing RFOT ideas in a meanfield limit appropriate for
an infinite dimensional system. Because of the dimensionality,
activated flow events become forbidden below a critical temperature
and strict ``replica symmetry breaking'' occurs so that the system can
remain forever trapped in one of an exponentially large number of
states. This $D=\infty$ meanfield analysis has led also to a rich set
of predictions that mirror many of the ideas about jamming that had
emerged from the Maxwell-inspired constraint theory of amorphous
assemblies.~\cite{LiuNagelSaarlosWyart} In the meanfield models, these
Maxwell constraints become just marginally satisfied at the so-called
Gardner transition, ~\cite{2015arXiv150107244C} which corresponds to
an isostatic arrangement of particles. While everything is
theoretically crisp in the mean field limit, how these ideas are to be
carried over to systems in finite dimensions with local interactions,
where finite time scales enter, is a question of kinetics not
thermodynamics. This topic is addressed in the present paper.

Within the RFOT theory, the emergence of stable states of repose out
of the equilibrium liquid appears as a type of
spinodal.~\cite{dens_F1, dens_F2} In addition, several lines of
argument suggest there is an isomorphism of the emergence of the
structural glass state with a spinodal of a random field
magnet.~\cite{KTW, stevenson:194505} At a spinodal, excitations are
fractal and resemble strings or percolation
clusters.~\cite{1984PhRvB..29.2698U} Using this idea, Stevenson,
Schmalian, and Wolynes~\cite{SSW} described how to locate the point on
the equilibrium phase diagram where activated dynamics would start to
take over from the collisional transport, the so called landscape
regime. At temperatures above this transition point, mode coupling
theory describes the slowing of the dynamics but the MCT singularities
are smoothed by the lower barrier activated events which become the
dominant relaxation mechanism at still lower temperatures.~\cite{MCT1,
  LW_soft, PhysRevE.72.031509} These fractal excitations also exist in
a non-equilibrium glass and there, they give rise to secondary or beta
relaxations.~\cite{JohariGoldstein} Interestingly Kirkpatrick and
Wolynes~\cite{MCT1} speculated that the Gardner transition in the
meanfield Potts glasses discussed by Gross et
al.~\cite{GrossKanterSomp} was connected to these secondary
relaxations in the structural glasses already in 1987. The approach we
pursue here is to see how these fractal excitations not only remain in
the glassy state but can themselves proliferate leading to mechanical
instability if an amorphous solid is put under high pressure. This
instability occurs because the rigidity of the amorphous matrix allows
it to maintain stresses like an arch or dome about a region that
reconfigures thereby adding to the driving force for such
reconfiguration events. The resulting analysis leads to a diagram that
highlights how low barrier processes can arise at high pressure in a
finite dimensional system near a jamming point. This diagram bears
some resemblance to the well-studied phase diagram of the infinite
dimensional hard sphere system but is distinct in several essential
aspects. Most importantly, the strict phase transition lines of the
$D=\infty$ system turn into dynamical glass transition lines that
depend on time scales of observation.

Owing to the presence of discrete spatial and translational
symmetries, the analysis of the mechanical stability of periodic
crystals can be reduced to computing, for each reciprocal lattice
vector, the frequencies of a finite number of vibrational modes that
are relatively straightforward to identify. For a broad class of
inter-particle potentials, such periodic structures can be seen to
remain stable just above absolute zero by computing the eigenvalue
spectrum of the Hessian of the potential energy
landscape.~\cite{Ashcroft} This analysis can be extended to highly
anharmonic systems---like hard spheres at finite temperatures---using
the density functional theory (DFT) or self-consistent phonon
theory.~\cite{Fixman1969} For a pedagogical review, see
Ref.~\onlinecite{L_AP}. The hard sphere crystal remains solid at
finite temperature even though the particles are virtually never in
direct contact except during collisions. Similar stability arguments
have been successfully applied to {\em aperiodic} collections of both
rigid and soft particles.~\cite{dens_F1, Lowen, RL_LJ} The landscapes
of those systems develop free energy minima corresponding to localized
particles below a crossover centered at a temperature $T_\scr$ (or
density $\rho_\scr$). In crucial distinction from perfect periodic
crystals, which are nearly unique, the number of aperiodic free energy
minima scales exponentially with the system size. The overlap of these
many basins of attraction leads to a new route to instability.

\begin{figure}[t]
  \includegraphics[width= 0.9 \figurewidth]{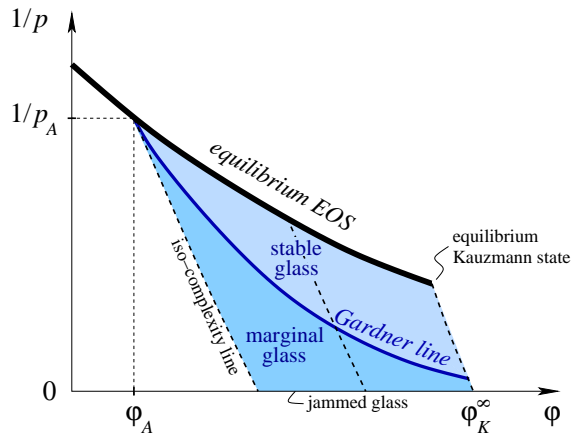}
  \caption{\label{MF} The phase diagram of a meanfield liquid in the
    $(\varphi, p^{-1})$ plane, after Ref.~\cite{Berthier26072016} The
    quantities $\varphi$ and $p$ are the filling fraction and pressure
    respectively. EOS~$=$~equation of state. Note reports on the
    location of the low-$\varphi$ end of the Gardner line
    vary.~\cite{Charbonneau2014}}
\end{figure}

The two views of the stability of aperiodic crystals---based on
Maxwell-like and landscape-derived notions respectively---were
reconciled in a meanfield model of glassy liquids by Mari, Krzakala,
and Kurchan~\cite{PhysRevLett.103.025701, doi:10.1063/1.3626802}
(MKK). A one-stage replica symmetry breaking (RSB) transition emerges
at a temperature $T_A$, at which the free energy landscape develops
many distinct yet equivalent minima. The number of metastable minima
scales exponentially with the system size. ($T_A$ is the meanfield
analog of the temperature $T_\scr$.~\cite{LW_soft}) Because in the
meanfield theory the minima are separated by infinite barriers, the
system must remain, upon further quenching, in a single one of these
minima of the free energy. If the system is then cooled further, the
pressure and density would follow a curve that coincides with the
lowest density iso-complexity line in Fig.~\ref{MF}. By definition,
the complexity is the logarithm of the multiplicity of the free energy
minima, per particle. Yet when Boltzmann-averaged over the totality of
the minima, the resulting equation of state is simply a smooth
continuation, toward higher densities, of the equilibrium equation of
state of the uniform liquid. As a formal construct, one may imagine an
ideal non-local move set that would allow one to continue to compress
the liquid along the so obtained ``equilibrium'' equation of state and
then perform a quench starting at any temperature below $T_A$. Such a
quench results in the usual glass of laboratory experience and, in the
infinite-pressure limit, a {\em jammed} configuration that had been
arrested in a specific aperiodic minimum, please see the graphical
illustration in Fig.~\ref{MF}.  The behavior of meanfield glassy
liquids is however both apparently subtler and richer. On further
cooling, the one-stage RSB is followed by a full---or
continuous---RSB, in which each individual minimum breaks up into an
infinite number of sub-basins with a distributed degree of
overlap.~\cite{doi:10.1021/jp402235d} This break-up is analogous to
the Gardner transitions in $p$-spin models.~\cite{Gardner1985747}
Simulations of the MKK model indicate this apparent symmetry breaking
is accompanied by a mechanical instability.~\cite{2015arXiv150107244C}
Whether this higher order transition will occur in finite dimensions
with local interactions is unclear.~\cite{PhysRevLett.114.015701}
Urbani and Biroli~\cite{PhysRevB.91.100202} have argued that full RSB
is either absent in spatial dimensions less than 6 or perhaps could be
conserved through a non-perturbative mechanism. Recent studies of soft
polydisperse spheres~\cite{2017arXiv170604112S} suggest that in finite
dimensions, quench-induced instabilities do not span the system but
are spatially localized and only sporadic; no susceptibilities seem to
diverge.

The appearance of marginally stable modes in the meanfield MKK model
and the $D=\infty$ hard sphere glass~\cite{PhysRevLett.103.025701,
  doi:10.1063/1.3626802} on approach to
jamming~\cite{LiuNagelSaarlosWyart} has been
proposed~\cite{Berthier26072016} as a possible explanation for
anomalies observed in cryogenic glasses, e.g., the Boson Peak and
two-level systems.~\cite{LowTProp, Esquinazi} Although suggestive,
theories based on this proposal have not yet produced quantitative
predictions. In particular, they do not explain the quantitative
universality seen in these low temperature
properties.~\cite{FreemanAnderson, YuLeggett} On the other hand,
arguments~\cite{LW_RMP, LW, LW_BP} within the free-energy landscape
framework already quantitatively account for those puzzling cryogenic
phenomena by quantizing the reconfigurational motions that normally
equilibrate the supercooled liquid. Only a fraction of the molecular
motions in question exhibit low enough barriers to be realized near
absolute zero quantum mechanically.

\begin{figure}[t]
  \includegraphics[width=  \figurewidth]{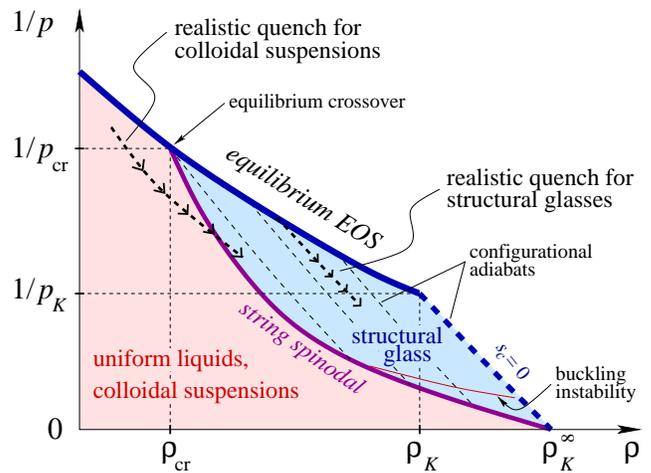}
  \caption{\label{nonMF} The off-equilibrium diagram of a
    non-meanfield liquid in the $(\varphi, \rho^{-1})$ plane, where
    $\rho$ is the number density. $s_c$ stands for the configurational
    entropy. Liquid states cannot exist to the right of the $s_c=0$
    line, by construction. Note that the ``string spinodal'' is a
    crossover, not a sharp boundary, in view of the limited spatial
    extent of aperiodic free energy minima in finite dimensions.  See
    Fig.~\ref{HS} for approximate evaluation of the sector boundaries
    for monodisperse hard spheres. }
\end{figure}

Here we treat jamming, aging, and the emergence of rigidity in a
unified fashion that goes beyond meanfield theory. The key is
accounting for the effects of local density changes during structural
relaxations. These density effects are shown to be very significant
for deep quenches or highly compressed systems. The results of our
analysis are graphically summarized in the diagram in
Fig.~\ref{nonMF}.  To understand this diagram we must realize that
relaxation times can be very long typically but are not strictly
infinite in systems with local interactions. Only the transition at
the possible entropy crisis where $s_c=0$ can be strictly considered a
true phase transition in this diagram. Many states with different
pressures and densities throughout the plane can be achieved depending
on the sample history. We consider all the states in the colored part
of the diagram as being potentially accessible. Even under uniform
cooling or compression, many different quenching protocols can be
realized depending on the speed of quenching. Obviously, many
different sequences of cooling and compression or heating and pressure
release can be envisioned.  Quenches that start below the crossover to
activated transport result in a frozen glass (the blue region on the
diagram) whose structure initially is not significantly
configurationally perturbed relative to the original equilibrated
liquid. Each particle largely maintains the same set of neighbors
which had formed a transient cage around it already in the equilibrium
state. The system falls out of equilibrium when the relaxation and
quenching time scales match.  Still, aging towards lower free energy
states will occur via local, compact activated processes, generally
accompanied by a significant increase in the density along with a
decrease in the pressure; these aging events can be quantified as we
have described in our previous work on aging.~\cite{LW_aging}
Following an activated aging event, the existing cages in the
aperiodic crystal are locally replaced by slightly different cages. In
the reconfiguring region, individual particles move distances
comparable to the typical vibrational displacement within a cage. Each
particle nevertheless remains surrounded by the same set of nearest
neighbors as before the event. Several reconfiguration events would be
needed for the identity of the neighbors to be fully reset. Save for
effects of volume relaxation, which we will now focus on, the above
microscopic picture is very much like that developed earlier by us
where we focused primarily of shallow thermal
quenches.~\cite{LW_aging} We see that amorphous solids can regarded as
mechanically stable at all lengthscales below a cooperativity size
$N^*$; this scale depends on the time scale of the quench. Amorphous
solids are only {\em metastable} on scales larger than $N^*$, which is
determined by an argument that resembles nucleation theory.

Suppose we continue a thermal quench. Once the quench becomes
sufficiently deep, the vibrational amplitude of a given particle may
become less than the gap to some of its neighbors so the particle
finds itself colliding with a smaller set of neighbors than it was
making contact with in the beginning of the quench. This breaking of
local symmetry comes from the individual cages not being strictly
isotropic when the system comes out of equilibrium. If the number of
long-lived collisional contacts becomes less than the number of
Maxwell constraints, the system becomes locally unstable.  The
microscopic consequences of the resulting instability strongly depend
both on the density and pressure. We emphasize that for systems away
from equilibrium, there is no one-to-one correspondence between
pressure and density (at a fixed temperature) Both parameters must be
specified separately in order to describe both instability and jamming
at high pressures.

We will see that for a sufficiently deep quench, the local relaxations
result in such a significant increase in coordination---and a
concomitant local pressure decrease---that the energy released in
reconfiguring can drive the barrier for reconfiguration to zero.
Particle in the reconfigured region will move a distance significantly
exceeding the typical vibrational amplitude thus destroying the
existing cages and then forming new cages using a distinct set of
neighbors, in an avalanche-like fashion.  When one particle escapes
its existing cage, the avalanche motion will resemble a string whose
direction changes.  The avalanche also can branch leading to a fractal
percolation-like cluster of rearranged particles.  This instability
toward fractal motions is well documented in equilibrated glassy
liquids:~\cite{Glotzer_strings} The mathematics resembles a
``string-deconfinement'' transition and has been described by
Stevenson, Schmalian, Wolynes.~\cite{SSW} When strings proliferate,
the motions change from being largely activated to
collisional~\cite{LW_soft} and can be described by extended
mode-coupling theory.  This transition is a crossover {\em
  spinodal}. We will also observe that for sufficiently rigid
particles, the string instability seems to be related to a set of
significantly milder instabilities in which particles confined to
anisotropic cages will have to choose one end of the cage over the
other. This other end of the cage will partially collapse---or
``buckle''---upon further quenching, resulting in a local jamming
event.

We note that achieving sufficiently high dimensionless pressures in
molecular systems, in order to encounter the transitions to marginal
stability, is hard. For these reasons, we believe quench-induced
marginal stability is unlikely to underlie the low temperature
anomalies observed in the laboratory for structural glasses.  On the
other hand, in granular and colloidal particles, these high
dimensionless pressures can be achieved but these systems are made of
larger constituents and are difficult to equilibrate at high densities
owing to their intrinsically long time scales for particle
motion. Colloidal assemblies are typically prepared at pressures and
densities where strings nucleate largely unimpeded.  Once frozen, such
systems will age largely by string-like motions, not compact
reconfigurations. We shall see that when such systems are densified
past the spinodal, the local increases in density lead rapidly to
activation barriers so the aging will be significantly slowed down. As
a result, colloidal assemblies will become spatially heterogeneous
collections of nearly mechanically-unstable configurations. Being
trapped in such configurations on experimentally relevant timescales
will lead to a nearly marginally stable, jammed structure at infinite
pressure. A quench starting from the low density phase and crossing
the string spinodal will rapidly lead to the onset of apparent
mechanical rigidity.

True criticality is not expected at the string spinodal or the
buckling instability in this high pressure region because at finite
temperature, long-range correlations will be destroyed by the
activated reconfigurations themselves just as the mode-coupling
singularity is short-cut by the activated reconfigurations to
equilibrium.~\cite{MCT1} The lack of such criticality excludes the
possibility of a true macroscopic instability---hence the article's
epigraph.  Nevertheless, long-range correlations may well be observed
on times less than the activated reconfiguration time.


\section{Reconfiguration barriers and the marginal stability in
  far-from-equilibrium glasses}

\begin{figure}[t]
  \includegraphics[width= \figurewidth]{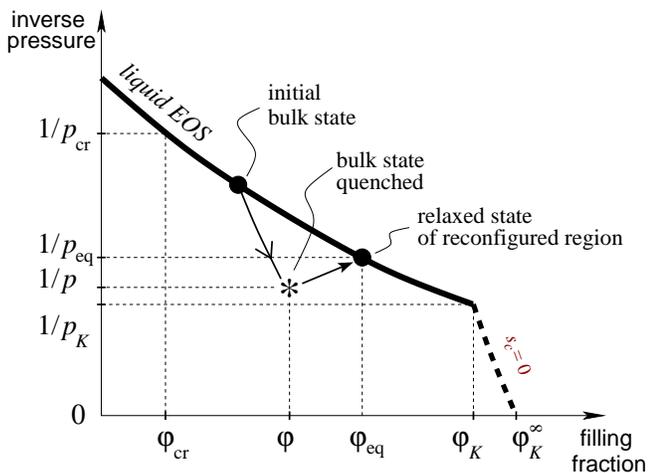}
  \caption{\label{protocol} Illustration of the liquid equation of
    state (EOS), a quenching protocol to a state $(\varphi, p)$, and
    the resulting state $(\varphi_\seq, p_\seq)$ of a compact
    reconfigured region.}
\end{figure}

It is useful to set the stage for the detailed analysis by reviewing
some basic notions about glassy behavior.  The equilibrium liquid
equation of state, which is graphically shown in Fig.~\ref{protocol}
by the thick line, ignores the possibility of crystallization to a
conventional periodic structure. (The ultimate fate of a supercooled
liquid is to crystallize.~\cite{SWultimateFate}) At a filling fraction
$\varphi_\scr$, the liquid nevertheless breaks translational symmetry
aperiodically. At a sufficient density, the free energy computed say
by density functional theory is no longer minimized by a spatially
uniform density profile.~\cite{dens_F1} Instead, a thermodynamically
optimal density profile can be well approximated by a superposition of
narrow Gaussian peaks centered at sites $\br_i$ of an aperiodic
lattice:~\cite{dens_F1, RL_Tcr}
\begin{equation} \label{rhoalpha} \rho(\br) = (\alpha/\pi)^{3/2}
  \sum_i e^{-\alpha(\br - \br_i)},
\end{equation}
The quantity $\alpha$ determines the magnitude of the thermal
fluctuations of particles localized at $\{ \br_i \}$. The lattice
specified by $\{ \br_i \}$ is associated with a mechanically stable
structure; the eigenvalues of the Hessian of the free energy
functional should be non-negative. Consequently, transitions to an
alternative aperiodic lattice must occur by activation.  In the
absence of fluctuations in the magnitude of the order parameter
$\alpha$, the onset of mechanical metastability with this ansatz
occurs at a sharply defined temperature $T_A$.  In finite dimensions,
the local order parameter can vary from site to site.  This allows
excitations that lead to a significant lowering of the onset
temperature of rigidity from the meanfield density functional value:
$T_\scr < T_A$ ($\rho_\scr > \rho_A$).~\cite{LW_soft, RL_LJ} Owing to
these excitations, the onset also becomes a gradual {\em
  crossover}.~\cite{KTW, XW, L_AP} The density $\rho_\scr$ is the
value of density at which the crossover to activated dynamics is
centered.

No thermodynamic quantities experience a singularity at $\rho_A$ in
the meanfield theory nor at $\rho_\scr$ in the finite-dimensional
analysis. This has been made clear by using the replica symmetry
breaking formalism.~\cite{MP_Wiley} At temperatures below $T_\scr$ the
aperiodic structure represented by $\{ \br_i \}$ persists locally for
long times compared to vibrational times but eventually transforms in
an activated fashion so that the liquid can flow. The corresponding
degrees of freedom labelling the distinct mechanically stable
configurations are usually referred to as ``configurational'' degrees
of freedom. Their thermodynamics is largely decoupled from the
vibrations, see Ref.~\onlinecite{L_AP} for a pedagogical exposition.
The key thermodynamic quantity is the configurational part of the
entropy, which reflects the multiplicity of the aperiodic free energy
minima.

To provide a concrete but, at best, semi-quantitative illustration, we
will use assemblies of monodisperse spheres as the model
liquid. Focusing on this system allows us also to make connection with
existing meanfield studies of hard spheres. It also allows us to find
explicit numerical estimates using approximate equations of state of
venerable vintage. For hard spheres, it is convenient to work with the
filling fraction $\varphi$:
\begin{equation} \varphi \equiv \rho (\pi \sigma_\text{hs}/6),
\end{equation}
where $\sigma_\text{hs}$ is the diameter of the sphere. Henceforth, we
will use the number density $\rho$ and the filling fraction $\varphi$
interchangeably. Real molecules are not hard spheres. Their
interactions are harshly repulsive but ultimately soft. At finite
temperature, we can consider molecules to have an effective core
diameter~\cite{LonguetHigginsWidom} that can be determined using an
unambiguous prescription such as those due to Barker and
Henderson~\cite{Hansen} or due to Weeks, Chandler, and
Anderson~\cite{WeeksChandlerAndersen} or their appropriate
generalizations for high density.~\cite{HallWolynes_JPCB} A key
quantity in our analysis will be the bulk modulus. The expression for
this quantity looks the same whether written in terms of $\rho$ or
$\varphi$:
\begin{equation} K \equiv - V \frac{\prtl p}{\prtl V} = \frac{\prtl
    p}{\prtl \ln \rho} = \frac{\prtl p}{\prtl \ln \varphi}.
\end{equation}

\begin{figure}[t]
  \includegraphics[width= \figurewidth]{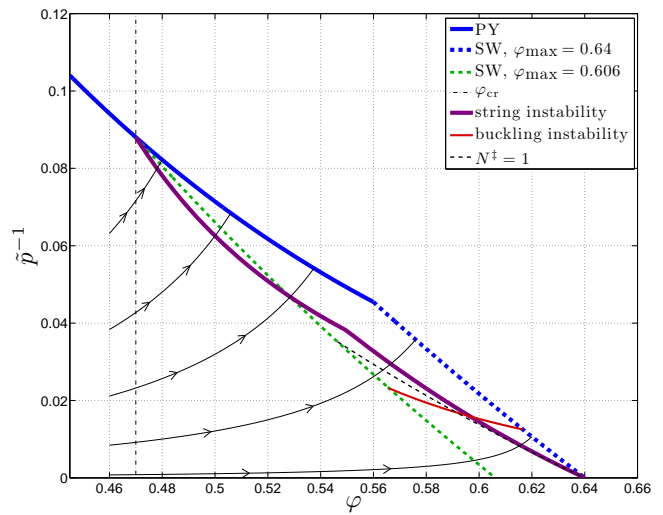}
  \caption{\label{HS} The off-equilibrium diagram for monodisperse
    hard spheres. The equilibrium equation of state (EOS) is
    approximated using the Percus-Yevick~\cite{McQuarrie} (PY)
    form. The Salsburg-Wood (SW) form from Eq.~(\ref{SW}) is used to
    approximate the configurational adiabats that cross the
    equilibrium EOS near the equilibrium crossover and the putative
    Kauzmann state.}
\end{figure}

After orienting ourselves by having looked at the quasi-equilibrium
curve, we now turn to discuss the glasses that result from quenches
that start at densities above $\varphi_\scr$. In this regime, suppose
a quench is made that allows for vibrational equilibration but is so
fast that no configurational equilibration whatsoever can occur. Each
particle would thus maintain the same set of neighbors as it had in
the beginning of the quench. In a meanfield sense, this neighbor set
would be described by a coordination number $z$. $z$ ultimately will
determine the maximum putative density that can be achieved by the
infinitely rapid quench.  The equation of state of a hard sphere
system, when confined to a single mechanically stable basin, is then,
approximately, $p \simeq k_B T/d a^2$.~\cite{RL_LJ} The quantity $d$
stands for the typical particle-particle gap in this basin and depends
on the coordination. The corresponding bulk modulus in a quenched
frozen state, $K \equiv - V(\prtl p/\prtl V)_T \simeq k_B T/3 d^2 a$,
thus exhibits a simple, ideal-gas like scaling relation with the
pressure:
\begin{equation} \label{Kp} \wtK \simeq A \wtp^2, \text{ if } \wtp
  \to \infty.
\end{equation}
In our analysis, it is convenient to use the dimensionless forms of
the two quantities:
\begin{align} \label{Kred}
  \wtK &\equiv K/\rho k_B T \\ \wtp &\equiv p/\rho k_B T.
\end{align}
The coefficient $A$ reflects the space dimensionality; in three
dimensions, $A=1/3$.  The relation is satisfied by the Salsburg-Wood
(SW) functional form:~\cite{doi:10.1063/1.1733163, HallWolynes_JPCB}
\begin{equation} \label{SW} \wtp = \frac{3}{\varphi_\smax/\varphi-1} +
  1,
\end{equation}
where $\varphi_\smax$ is a maximal putative density achievable by
rapidly quenching the starting structure; $\varphi_\smax$ depends on
the coordination in the original frozen state. It is a function of the
density and temperature where the system first fell out of
equilibrium.~\cite{Tool, Narayanaswamy, Moynihan} $\varphi_\smax$ can
be considered as providing equivalent information to the coordination
number $z$ in theories based on counting Maxwell constraints.  Since
reconfigurational motions are not allowed when making an infinitely
rapid quench the corresponding configurational heat capacity is zero,
so the configurational entropy remains constant during the quench:
$s_c = \text{const}$. The above protocol is approximately followed
during routine thermal quenches in the laboratory even though some
aging is expected to occur, if one waits long enough after first
falling out of equilibrium. We discussed the inhomogeneous structure
that occurs after waiting in our earlier paper.~\cite{LW_aging} Now,
there is a putative density, $\varphi_K$, above which $s_c = 0$, in
equilibrium. The third law makes this a fixed lower bound on the glass
transition temperature,~\cite{Kauzmann} which can only be reached for
infinitely slow cooling. This ideal, ``Kauzmann'' state corresponds
with the most stable aperiodic crystal and exists for liquids in
$D=\infty$.~\cite{mezard:1076} We show the $s_c = 0$ line in
Fig.~\ref{protocol} as the putative limiting case for the slowest
possible quench.
At any rate, the vibrational part of the compressibility for particles
depends on the particle-particle gaps but does not depend very much on
the precise coordination pattern. The slopes of the $s_c =
\text{const}$ lines, shown by thin dashed lines on Fig.~\ref{nonMF},
are thus numerically similar for different values of $s_c$. The region
putatively accessible by such ``conventional'' quenches is depicted in
blue in the latter Figure. Configurational adiabats for monodisperse
hard spheres, as approximated using the Salsburg-Wood form, are shown
in Fig.~\ref{HS}.

We now analyze the stability of the amorphous solid and the spatial
extent and kinetics of structural relaxation of the amorphous states
that were obtained by quenching an equilibrated liquid with $\varphi >
\varphi_\scr$.
Because a rapidly quenched state is off-equilibrium, any further
relaxation technically corresponds with {\em aging}.  Conversely, if
we are still to be confined to a single basin, quenches must be
extraordinarily fast. Still, this means that the Salsburg-Wood
adiabats for relatively low values of $\varphi_\smax$ can be formally
defined and accessed.

Here we present a quantitative framework to describe aging that
generalizes the one that was developed earlier by us~\cite{LW_aging}
and extended by Wolynes and coworkers to include for the possibility
of achieving states of marginal stability.~\cite{SSW, SWbeta,
  Wisitsorasak02102012} We first review those results that assumed
that individual, aging-induced displacements do not significantly
exceed the typical vibrational amplitude. We do this first for
reconfigurations that are relatively compact and then generalize to
the stringy or fractal case. The free energy cost for a compact region
to rearrange, in a fixed environment, is well approximated by a sum of
a bulk contribution $\Delta g N$ and a mismatch penalty $\gamma
N^{1/2}$:~\cite{KTW, XW, LW_aging}
\begin{equation} \label{FN1} F(N) = \Delta g N + \gamma N^{1/2},
\end{equation}
where $N$ is the number of reconfigured particles. The square-root
scaling of the mismatch penalty is slower, at large $N$, than the
conventional surface scaling $N^{2/3}$. This non-classical scaling of
the mismatch between two aperiodic free energy minima comes about
because the strained region at the border between the displaced
particles and the original environment can typically lower its energy
by as much as $\sim N^{1/2}$ by locally replacing any region with an
equivalent structure chosen out of the random ensemble.~\cite{KTW, XW,
  LRactivated, CL_LG} This effects parallels the wetting effect
discovered by Villain for the random field Ising
magnet.~\cite{Villain} This free energy lowering naturally scales with
the magnitude of Gibbs free energy fluctuations~\cite{KTW,
  LRactivated} and is thus approximately determined by the bulk
modulus:~\cite{LRactivated, L_AP}
\begin{equation} \label{gamma2} \wtgm \equiv \gamma/k_B T =
  \wtK^{1/2}.
\end{equation}
There are other, numerically smaller contributions to the mismatch
penalty,~\cite{LRactivated} which we neglect here. In any event, the
mismatch penalty between two random structures---the originally
quenched structure and the structure it can locally relax to---must be
bounded from above by the bulk modulus of the stiffer of the two
structures. A better approximation is obtained by noting that all
lattice properties depend continuously on the coordinates~\cite{CL_LG}
and so from here on, we will assume the mismatch penalty is determined
by the value of the bulk modulus {\em at} the interface.

\begin{figure}[t]
  \includegraphics[width= .9 \figurewidth]{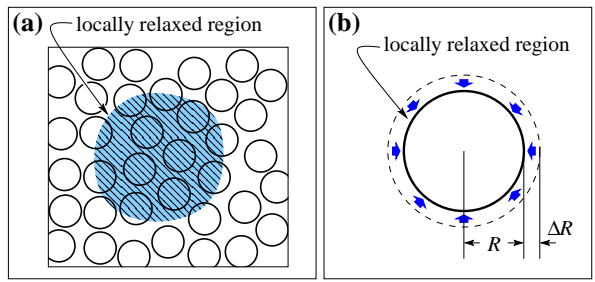}
  \caption{\label{relax} Cartoon of an aged region, in which the
    pressure and specific volume have decreased as a result of a
    reconfiguration event. Note the coordination in the rearranged
    central region will be larger than that in the bulk.}
\end{figure}

Consider now a quench toward the state denoted by the asterisk on
Fig.~\ref{protocol}.  The external pressure $p$ is higher than what it
would be in equilibrium, at the density and temperature in question,
because the structure we start with is more open than it would be in
macroscopic equilibrium. The reconfiguring region will thus relax to a
state having a lower specific volume---in which coordination is
higher---and equilibrate to some new value of pressure, call it
$p_\seq$, which is less than the external pressure, $p_\seq < p$. The
mechanically equilibrated pressure in the reconfigured region and its
immediate surrounding will be lower than the external pressure because
the environment is rigid!  The internal pressure can only {\em
  partially} accommodate for the contraction of the reconfigured
region since the shear modulus $\mu$ of the environment is finite.
These notions are graphically illustrated in Fig.~\ref{relax}.

The residual pressure mismatch is only modest for shallow
quenches.~\cite{L_AP} We therefore neglected this mismatch in our
earlier analysis of aging.~\cite{LW_aging} Here, in contrast, we want
to estimate the relaxation of pressure inside the reconfigured region
for arbitrarily high values of the external pressure $p$; such high
pressures are needed to achieve jamming.  To calculate the extent of
density relaxation, first suppose the volume of the relaxed region
will change by $\Delta V$ following an aging event. The relation
between the relative volume change, or packing fraction change,
$\Delta V/V = - \Delta \varphi/\varphi$ and the pressure drop can be
estimated using continuum, linear elasticity theory. This has been
described by Landau and Lifshits in the solved Problem 7.2 from
Ref.~\onlinecite{LLelast}, see also the Appendix of
Ref.~\onlinecite{L_AP}. We briefly reprise the argument here. For a
spherically-symmetric geometry, the deformation $\bu$ outside the
reconfigured region is directed along the radius-vector while its
magnitude is given by~\cite{LLelast} $u(r) = ar + b/r^2$. The linear
term, $ar$, describes the spatially uniform deformation caused by the
externally applied hydrostatic pressure, while the second term,
$b/r^2$, takes care of spatial inhomogeneities near the inclusion, if
any. Note that the deformation in the region's environment is pure
shear even though the {\em total} volume of the sample will decrease
as a result of the reconfiguration event by a fraction of $\Delta
V$.~\cite{L_AP} The resulting radial stress is given by the
expression~\cite{LLelast} $\sigma_{rr} (r) = 3K a - 4\mu b/r^3$.  The
coefficients $a$ and $b$ are fixed by the boundary conditions at the
interface of the inclusion $\sigma_{rr} (R)= - p_\seq$ and spatial
infinity $\sigma_{rr}(\infty)=-p$, yielding $b = - (p-p_\seq)
R^3/4\mu$. Thus the displacement of the boundary of the reconfigured
region, relative to its position before reconfiguration in the
presence of external pressure $p$, is given by $\Delta R = u(R) - aR =
- R(p - p_\seq)/4\mu$. In terms of the relative volume change $-\Delta
\varphi/\varphi = \Delta V/V = 3 \Delta R/R$ of the reconfigured
region, this yields for the pressure mismatch between the equilibrated
region and the bulk of the sample:
\begin{equation} \label{Deltaphi} p - p_\seq = \frac{4}{3} \,
  \frac{\Delta \varphi}{\varphi} \mu_\seq.
\end{equation}
Since the equation is correct up to terms of order $\Delta \varphi^2$,
$\varphi$ on the r.h.s. can stand for the density of either the
initial or final state; we will set $\varphi = \varphi_\seq$ for
convenience.  According to Eq.~(\ref{Kp}), the elasticity is clearly
non-linear in the high pressure limit. Still, Eq.~(\ref{Deltaphi})
indicates that most of the deformation will take place in the softer
regions, where the elastic constants are smallest: $\Delta \varphi
\propto 1/\mu$. Accordingly, we use in Eq.~(\ref{Deltaphi}) the value
of the shear modulus pertaining to pressure $p_\seq$ as pertinent to
the region just outside the reconfigured region. The label ``eq'' at
$\mu_\seq$ in the r.h.s. explicitly refers to the fact. In any event,
Eq.~(\ref{Deltaphi}) becomes more accurate the smaller $\Delta
\varphi$ is. Now, the value of pressure in the relaxed region is
determined by simultaneous solution of Eq.~(\ref{Deltaphi}) and the
equilibrium equation of state from Fig.~\ref{protocol}. At lower
pressures (higher temperatures) one follows the $s_c > 0$ branch, but
at higher pressures (lower temperatures) one follows the ``Kauzmann''
portion $s_c = 0$:
\begin{equation} \label{EOS} p_\seq = p_\text{\tiny
    eEOS}(\varphi_\seq),
\end{equation}
where the label ``eEOS'' signifies ``equilibrium equation of state''
It is not difficult to convince oneself that such a solution to the
nonlinear elastic balance always exists, as is explicitly illustrated
in Fig.~\ref{HS} for hard spheres: As a practical matter, it is most
convenient to first find and plot the sets of quenched states
$(\varphi, p)$ that would relax locally into a particular equilibrated
state $(\varphi_\seq, p_\seq)$, because the equilibrium EOS has a
discontinuity in slope at the Kauzmann state $(\varphi_K,
p_K)$. Examples of such paths of relaxation are shown in Fig.~\ref{HS}
as curves with arrows. By the collapse of the interior of a
reconfigured region, any initial state belonging to such a curve would
relax, within that region, to the equilibrium state located at the
high density end of the curve.  To avoid confusion we remind the
reader that the intermediate portions of these ``local-collapse
curves'' do not correspond to kinetic intermediates; a single aging
event leads the system directly to the state on the equilibrium
equation-of-state line in the interior of the aged region. To solve
for pairs $(\varphi, p)$, $(\varphi_\seq, p_\seq)$ of quenched and
relaxed states, respectively, we have expressed the shear modulus
through the bulk modulus and the Poisson ratio $\sigma =
(3K-2\mu)/2(3K+\mu)$:
\begin{equation} \label{muK} \mu = \frac{3(1-2\sigma)}{2(1+\sigma)}
  K_{s_c}.
\end{equation}
For concreteness, we have also assumed a value of the Poisson ratio
characteristic of the aperiodic crystal made of hard
spheres,~\cite{Lowen} see the Supporting Information for details. It
is essential that the bulk modulus used in Eqs.~(\ref{Deltaphi}) and
(\ref{muK}) be not the isothermal kind $(\prtl p/\prtl \ln
\varphi)_T$, but the one computed along the appropriate $s_c =
\text{const}$ lines, as is emphasized by the subscript $s_c$ in
Eq.~(\ref{muK}). This is, again, because the surrounding matrix is
assumed to relax exclusively by vibrational readjustments.

Because the overall sample is maintained at constant external pressure
and temperature, the relevant driving force in Eq.~(\ref{FN1}) is the
Gibbs free energy.~\cite{LLstat} This can be seen by writing down the
expression for the probability for a system to have energy $E$ and
volume $V$ when subject to an external pressure $p$ and temperature
$T$: $p(E, V) \propto e^{-\beta (E - T S + pV)}$. The total entropy
includes both a vibrational and configurational contribution. Since
aging-induced escape occurs from a particular aperiodic minimum, the
configurational entropy of the initial state is zero by
construction. Thus in the usual way,~\cite{LW_aging, LRactivated} the
driving force for reconfiguration is given by the change in the Gibbs
free energy of an individual aperiodic state and an entropic
contribution due to the multiplicity of the target state.  Per
particle, it is given by the quantity
\begin{equation} \label{dg} \Delta g = p_\seq/\rho_\seq - p/\rho - T
  s_c(p_\seq) - T \Delta s_\svibr < 0.
\end{equation}
The first two terms on the r.h.s. account for the enthalpic
stabilization of the region due to the pressure drop.  This is nonzero
because the environment is a rigid solid, not a stress-free fluid, as
already discussed. The $- T s_c(p_\seq)$ terms accounts for the
multiplicity of the free energy minima in the equilibrium state, and
the term $- T \Delta s_\svibr$ corresponds with the stabilization due
to the increase in the vibrational entropy upon reconfiguration. The
$\Delta s_\svibr$ term stems from the increased spacing $d$ between
the particles in a state with a higher coordination number. Since for
rigid objects $d \propto 1/p$, one can approximately write:
\begin{equation} \label{dsvibr} \Delta s_\svibr = 3 \ln(p/p_\seq).
\end{equation}
For soft particles, the dependence of the vibrational amplitude---and
thus vibrational entropy---on pressure is more complicated and,
furthermore, system-dependent. In any event, this dependence is
significantly weaker than the already mild, logarithmic dependence in
Eq.~(\ref{dsvibr}). There is another potential contribution to the
driving force in Eq.~(\ref{dg}), viz., any vibrational energy
difference $\Delta E_\svibr$ between the quenched and target state of
the reconfigured region, which would generally depend on anharmonicity
and quantum effects.  For rigid objects, this difference is strictly
zero. For classical oscillators, it also vanishes in the harmonic
limit. Accounting for effects of anharmonicity would not significantly
improve the accuracy of approximation, see the discussion of
Eq.~(\ref{Deltaphi}); these effects, if any, will be therefore
ignored.
Finally, we inquire whether there is an enthalpic stabilization of the
{\em environment} of the reconfigured region. For linear elasticity,
the environment experiences only shear but no uniform dilation, as
already mentioned.  Thus the radial stress is exactly compensated by
the lateral stress and so, no stabilization is expected at this level
of approximation. The next order term will scale with $\Delta \varphi$
or higher and will be ignored henceforth. To avoid confusion we note
that the free energy cost of creating an interface between the
environment and the reconfigured region has been already included
through the mismatch penalty $\gamma N^{1/2}$.

Now, the critical size $N^\ddagger$ at which the free energy profile
(\ref{FN1}) for compact reconfiguration events reaches its
saddle-point value is easily computed:
\begin{equation} \label{Ncrit} N^\ddagger = \left(
    \frac{\gamma}{2 \Delta g} \right)^2,
\end{equation}
while the activation barrier itself equals:
\begin{equation} \label{Falpha}
F^\ddagger_\alpha = \gamma^2/4(-\Delta g).
\end{equation}
The corresponding relaxation time is
\begin{equation} \label{tau} \tau = \tau_0 e^{F^\ddagger_\alpha/k_B
    T}.
\end{equation}
We remind the reader that in the equilibrium fluid regime, this
barrier reproduces the venerable Vogel-Fulcher law. For shallow
quenches, it gives the characteristic ``non-linearity'' of a changed
Arrhenius relaxation during aging.~\cite{LW_aging} Together with
Eq.~(\ref{dg}), the above result for $F^\ddagger$ indicates that
structural relaxation events in samples quenched by compression occur
more rapidly than they would in samples equilibrated at pressure
$p_\seq$, where the only driving force is entropic: $\Delta g = - T
s_c$. This is analogous to the way a sample being slowly {\em
  thermally} quenched starting at temperature $T_1$ to the ambient
temperature $T_2 < T_1$, has an extra driving force for relaxation due
to the enthalpic stabilization~\cite{LW_aging} $\int_{T_1}^{T_2}
\Delta C_p dT < 0$. Here, $ \Delta C_p$ stands for the configurational
specific heat. Any externally imposed shear also would add to the
driving force ultimately giving the yield strength of the
glass.~\cite{Wisitsorasak02102012}

The cooperativity size $N^*$, defined by the relation $F(N^*)=0$:
\begin{equation} \label{N*} N^* = \left( \frac{\gamma}{\Delta g}
  \right)^2,
\end{equation}
indicates the size of a region that is guaranteed to have another
aperiodic mechanically stable state.  The quantity $N^*$ is very
important because it signifies the size below which the sample is {\em
  mechanically stable}. In contrast, a larger region is only
metastable on the time scale defined by the barrier from
Eq.~(\ref{Falpha}). Note the simple relation 
\begin{equation} \label{NN}
N^* = 4N^\ddagger.
\end{equation}

\begin{figure}[t]
  \includegraphics[width= \figurewidth]{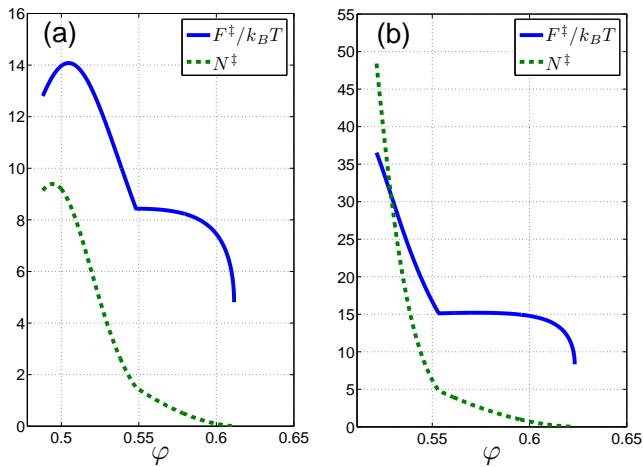}
  \caption{\label{brN} Density dependence of the dimensionless
    activation barrier and critical size for structures obtained by
    quenching along Salsburg-Wood (SW) adiabats that intersect the
    equilibrium EOS at $\varphi = 0.4880$ ($\varphi_\smax = 0.6114$)
    in panel {(a)}, and $\varphi = 0.5213$ ($\varphi_\smax = 0.6232$)
    in panel {(b)}. }
\end{figure}

How do the barrier and critical size for an aging reconfiguration
event depend on the depth of the quench?  In Fig.~\ref{brN}, we show
these two quantities as functions of the filling fraction in the
quenched structure. Each panel of the figure exclusively pertains to
barriers found from states obtained by quenching along a specific
configurational adiabat, $s_c = \text{const}$, approximated according
to Eq.~(\ref{SW}). The specific value of $\varphi_\smax$ chosen in
panel (a) refers to the Salsburg-Wood adiabat that is near but not too
close to the spinodal crossover so as to avoid barrier softening
effects due to spatial variations of the order parameter
$\alpha$.~\cite{LW_soft} The value of $\varphi_\smax$ chosen for
plotting in panel (b) corresponds with a barrier at equilibrium that
is numerically close to the typical barrier at the laboratory glass
transition, i.e., $F^\ddagger/k_B T = \ln(10^4 \text{sec}/10^{-12}
\text{sec}) \approx 37$. Incidentally, we observe that the value of
the critical size $N^\ddagger$ found with these equation of states is
in remarkably good agreement with earlier predictions of the RFOT
theory.~\cite{XW, LW, RWLbarrier, LRactivated} Here and everywhere
below, we parametrize the density dependence of the configurational
entropy, in equilibrium, using an interpolation formula $s_c(\varphi)
= s_c(\varphi_\scr) \, (\varphi_K - \varphi)/(\varphi_K -
\varphi_\scr)$ with $s_c(\varphi_\scr) = 1.75 k_B$.~\cite{RL_LJ,
  RL_Tcr} We assume the configurational part of the entropy remains
constant along the Salsburg-Wood adiabats, as mentioned.

We notice that for sufficiently high pressures, the barrier eventually
decreases with pressure and saturates at a value that depends on the
proximity of the Salsburg-Wood adiabat to the Kauzmann line, at which
by construction $\varphi_\smax = \varphi_K^\infty$:
\begin{equation} \label{FT} \lim_{p \to \infty} \frac{F^\ddagger}{k_B
    T} = \frac{\varphi_\smax}{\varphi_K^\infty - \varphi_\smax} \:
  \frac{1+\sigma}{8(1-2\sigma)},
\end{equation}
see the Supplementary Information. Here, $\varphi_\smax$ pertains to
the quench in question. The low-pressure behavior of the barriers, on
the other hand, depends on the proximity of the Salsburg-Wood adiabats
to the equilibrium crossover. For sufficiently small $\varphi_\smax$
and low pressures, the barrier at first {\em increases} with the depth
of the quench before adopting the aforementioned high-pressure
trend. The increase of the thermally scaled barrier with the extent of
quench is, in fact, observed in routine thermal quenches in the
laboratory, consistent with the expectation from this
calculation.~\cite{LW_soft}

It is instructive to note that in the high pressure limit, we can
combine Eqs.~(\ref{Deltaphi})-(\ref{muK}) with (\ref{Kp}) to obtain:
\begin{equation} \label{dpphigh} \varphi_\smax \wtp - \varphi_K^\infty
  \wtp_\seq = \wtp_\seq^2 \: (\varphi_K^\infty -
  \varphi_\smax) \: \frac{2(1-2\sigma)}{3(1+\sigma)}.
\end{equation}
This result implies that the pressure inside the reconfigured region
scales only as the square root of the external pressure:
\begin{equation} p_\seq = p^{1/2} \left[ \frac{\varphi_\smax}{
      (\varphi_K^\infty - \varphi_\smax)}
    \frac{3(1+\sigma)}{2(1-2\sigma)} \right]^{1/2}, \hspace{3mm} p \to
  \infty
\end{equation}
and thus the relaxed inerior pressure becomes arbitrarily smaller than
than the external pressure as the latter grows to be infinite. The
resulting, arbitrarily large pressure discontinuity arising from the
matrix' rigidity implies that the outside region is unable to relax to
fill the effective void formed within the reconfigured region and,
thus, is {\em jammed} in a dome-like configuration. While domes made
of engineered blocks can be perfectly stable, no such stability can be
expected from a dome made of frictionless spheres with finite thermal
vibration. At the same time, we have seen that while the
reconfiguration barrier saturates at a finite value, with increasing
pressure, the inter-particle forces can be made arbitrarily large, in
which case {\em finite}-sized displacements of particle become
virtually barrierless: If formed in this regime, the ``dome'' would
promptly collapse from any thermal motion! Another way to look at this
is that the vibrational displacement in the original quenched
structure can be made arbitrarily small by sufficient external
compression. As a result of the pressure mismatch, $(p - p_\seq)$, the
aging-induced displacement of the particle can then exceed this
typical vibrational displacement to an arbitrarily large degree
implying that the vibrational cage is destroyed by thermal motion.

\begin{figure}[t]
  \includegraphics[width= .9 \figurewidth]{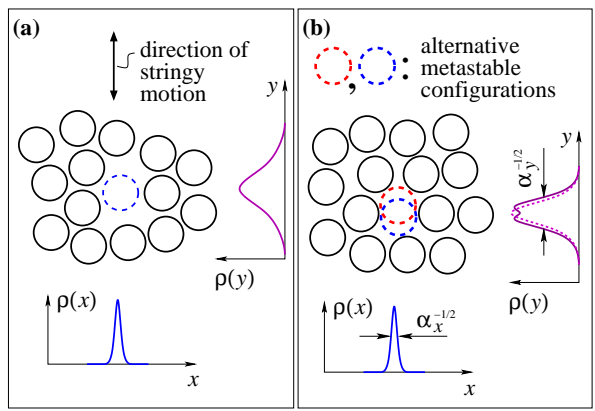}
  \caption{\label{jamstrings} {\bf (a)} A stringy instability is
    signalled by the ability of a particle to escape the cage via a
    displacement that significantly exceeds the typical vibrational
    displacement. The vibrational amplitude in the direction of lower
    pressure exceeds its typical value, while the collision frequency
    is lowered.  {\bf (b)} Illustration of how compression of an
    anisotropic cage can lead to a local symmetry breaking upon which
    the caged particle will be subject to a bistable effective
    potential and will have to choose one of the minima upon further
    compression, while leaving the remainder of the cage slightly
    undercoordinated. }
\end{figure}

We see that the critical size $N^\ddagger$ for compact rearrangements
could become as small as one wishes, apparently, even smaller than
one. (For the reader's reference, Fig.~\ref{HS} shows where, on the
Salsburg-Wod adiabats, the critical size becomes exactly equal to 1.)
This means there will be individual one-particle reconfigurations that
have small barriers. Thus a series of reconfiguration events would
then occur in a spatially contiguous fashion so as to form {\em
  strings} and, ultimately, percolation clusters, see
Fig.~\ref{jamstrings}(a). Indeed this means the instability threshold
itself will occur with less additional driving force than we have
computed assuming compact reconfiguration events. Stevenson,
Schmalian, and Wolynes~\cite{SSW} determined the free energy cost for
forming such strings or fractal ``lattice animals'' when in the liquid
state, i.e., for equilibrium conditions. Their analysis was later
generalized by Wisitsorasak and Wolynes to accommodate for the
presence of additional bulk driving forces as can occur in the glass,
such as mechanical stress.~\cite{Wisitsorasak02102012} The free energy
profile for creating a string or fractal cluster, like the compact
rearrangement, can be written as a function of size $N$ but contains
an additional size-dependent shape entropy of the reconfigured region:
\begin{equation} \label{FN3} F(N) = \gamma' N + \Delta g N - k_B T
  (\ln \Omega) N,
\end{equation}
Here $\gamma'$ is the extrapolated free energy cost to reconfigure one
particle. Similarly to the coefficient $\gamma$ from
Eq.~(\ref{FN1}). The quantity $\gamma'$ reflects the cost of breaking
nearly all the contacts around the particle and is a fraction of the
mismatch coefficient $\gamma$ for compact clusters:~\cite{SSW}
\begin{equation} \label{cgamma} c_\gamma \equiv \gamma'/\gamma < 1.
\end{equation}
We expect the above ratio not to be singular and remain roughly
constant.  The linear size-dependence of the penalty for forming a
string or fractal cluster of size $N$, $\gamma' N$, follows from
geometry.  The quantity $\Omega$ reflects the multiplicity of strings
that can emanate from a given locale; its value can be adjusted to
effectively account for string percolation using known properties of
percolation clusters.~\cite{SSW} Numerically, one has $k_B \ln \Omega
\approx 2 k_B$, a value we adopt here for concreteness, also gives the
correct scaling of the crossover temperature with
fragility.~\cite{SSW}

According to Eq.~(\ref{FN3}), fractal excitations can form near a
spinodal once the slope of the $N$ dependence is non-positive:
\begin{equation} \label{ineq} (\gamma' + \Delta g - k_B T \ln \Omega)
  \le 0.
\end{equation}
For equilibrium states, the equality corresponds with the crossover
density $\varphi_\scr$ in the liquid.  Below the latter density, the
strings are technically infinite in length, corresponding with the
uniform liquid, while for $\varphi > \varphi_\scr$ the strings are
only finite and a compact region must rearrange for flow.~\cite{SSW}

In quenched states, we must determine where on the arrowed curves in
Fig.~\ref{HS} the condition (\ref{ineq}) for marginal stability is
satisfied since the driving force depends on the amount of
contraction.  It follows that the highest density at which the
condition (\ref{ineq}) is still satisfied is given by the equation:
\begin{equation} \label{cond1} \wtgm' = \wtp^\text{(s)}-\wtp_\seq +
  s_c/k_B + \ln \Omega + \Delta s_\svibr,
\end{equation}
where we used Eqs.~(\ref{dg}) and (\ref{dsvibr}). The label ``(s)''
signifies ``spinodal,'' in reference to the point of marginal
stability.

As before, pairs of a quenched state and a corresponding relaxed state
are determined by simultaneous solution of Eqs.~(\ref{Deltaphi}) and
(\ref{EOS}).  The loci $\wtp^\text{(s)}(\varphi)$, that result from
the calculation for hard spheres, are plotted as the ``string
spinodal'' line in Fig.~\ref{HS}.  Note that at the {\em equilibrium}
crossover, where there is only an entropic force, the aforementioned
parametrizations $s_c(\varphi_\scr) = 1.75 k_B$ and $k_B \ln \Omega
\approx 2 k_B$ for our equations of state determine the appropriate
numerical value of $c_\gamma$, which takes on a reasonable value of
$0.49$.

As suggested by Fig.~\ref{HS}, the high-pressure end of the
string-spinodal line asymptotes to the infinite-pressure states that
are quenched from the Kauzmann point. This is indeed the case; the
corresponding scaling form can be straightforwardly determined and are
not affected by the precise parametrization of $s_c$, $\ln \Omega$,
and $\Delta s_\svibr$ but only the high-pressure asymptotics of the
shear modulus, see the Supplementary Information for explicit
expressions. The Supplementary Information also provides an expression
for the slope of the string-spinodal line, near the equilibrium
crossover, that uses experimentally measurable materials
constants.~\cite{RL_LJ, Lowen, RL_Tcr, 0305-4608-10-7-009,
  PhysRevB.80.132104} Our estimates indicate that just as for hard
spheres, the low pressure end of the spinodal line in molecular
substances should have a steeper slope, in the $(\varphi, p^{-1})$
plane, than the equation of state for a configurationally adiabatic
quench starting at the equilibrium crossover. We remind the reader
that owing to fluctuation and the emergence of activated dynamics, the
spinodal line signifies a gradual crossover, not a sharp boundary. The
line also represents a glass transition since on its high-density
side, the dynamics are activated and depend on the time scale of
observation and preparation.

\begin{figure}[t]
  \includegraphics[width= \figurewidth]{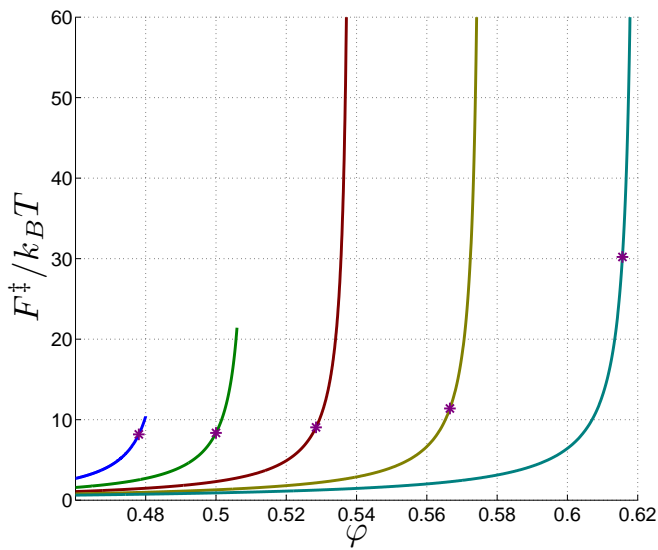}
  \caption{\label{pathBarrier} Density dependence of the dimensionless
    activation barrier from Eq.~(\ref{Falpha}) along the five
    local-collapse paths indicated by the arrowed curves in
    Fig.~\ref{HS}. The asterisks denote the predicted location of the
    spinodal from the string theory argument.  The expression
    Eq.~(\ref{Falpha}) does not include the barrier-softening effects
    near the spinodal.~\cite{LW_aging} In a more complete calculation,
    the activation barrier would vanish at the densities indicated by
    the asterisks.}
\end{figure}

\begin{figure}[t]
  \includegraphics[width= \figurewidth]{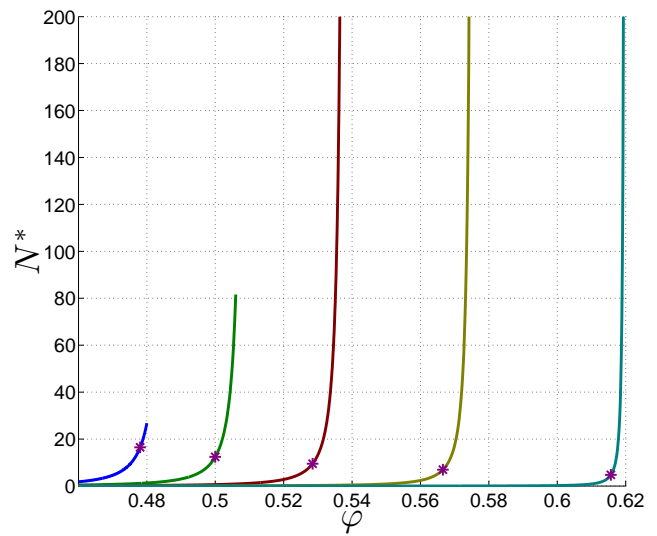}
  \caption{\label{pathN} Density dependence of the cooperativity size
    $N^*$ of compact clusters from Eq.~(\ref{N*}) along the five
    local-collapse paths indicated by the arrowed curves in
    Fig.~\ref{HS}. The asterisks denote the location of the string
    spinodal. Again, no softening effects are
    included.~\cite{LW_aging} If softening were included, the
    appropriate cooperativity size near the spinodal would correspond
    to the typical string length and would diverge except for string
    overlap effects.}
\end{figure}

\begin{figure}[t]
  \includegraphics[width= \figurewidth]{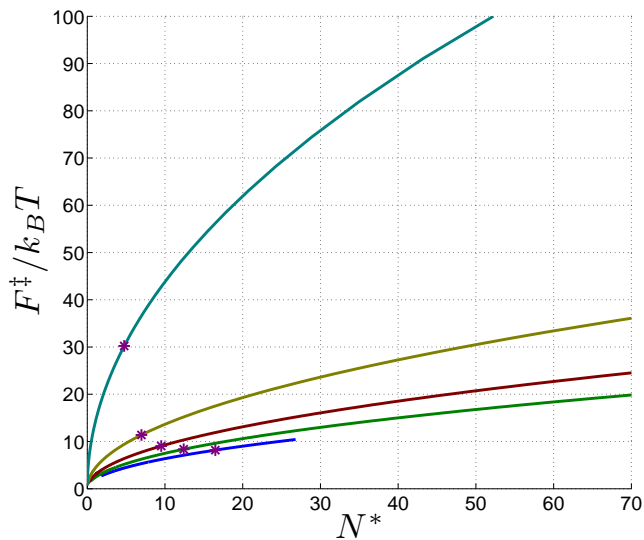}
  \caption{\label{pathNBr} Parametric plot of the dimensionless
    barrier $F^\ddagger/k_BT$ vs. the compact cluster cooperativity
    size $N^*$ from Figs.~\ref{pathBarrier} and \ref{pathN} along the
    five local-collapse paths indicated by the arrowed curves in
    Fig.~\ref{HS}. The asterisks denote the location of the predicted
    string spinodal. }
\end{figure}

The spinodal appears as a mechanical instability of the apparently
rigid state when coming from the {\em high} density side to lower
densities. String-mediated aging events locally destroy the existing
cages in an avalanche-like fashion. When approached from the
low-density, replica-symmetric side, however, the same line can be
viewed as signalling the onset of effective rigidity percolation, as
pointed out in the Introduction. After the spinodal line is crossed,
activation barriers begin to rise rapidly and the system typically
will fall out of equilibrium on a fixed observation time scale. The
density dependence of these barriers, for the five specific aging
paths from Fig.~\ref{HS}, are shown in Fig.~\ref{pathBarrier}. The
corresponding cooperativity sizes $N^*$, Eq.~(\ref{N*}), are shown in
Fig.~\ref{pathN}.  For the reader's reference, we have indicated the
location of the string instability, for each respective curve, with a
purple asterisk. In a complete treatment, one should include barrier
softening effects, which would then show both the computed barrier and
cooperativity size vanishing in a critical fashion at the
spinodal. The softening effects stem from spatial variations of the
local order parameters and effectively erase the activated barrier at
the spinodal, so the liquid motions are slowed exclusively owing to
mode-coupling effects or other effects such as the viscous drag of
colloidal particles against the solvent. Away from the spinodal, the
barrier-softening effects become increasingly less important, see
quantitative estimates for specific substances in
Ref.~\onlinecite{LW_aging}. In any event, a parametric plot of the
activation barrier $F^\ddagger$ vs. the cooperativity size of compact
clusters $N^*$ is expected to be insensitive to softening
effects. Such a parametric plot, shown in Fig.~\ref{pathNBr},
indicates that the onset of activation, upon crossing the spinodal
line from the low-density side, gradually changes in character with
increasing pressure. At a fixed value of the activation
barrier---which corresponds to a fixed observation time up to the
prefactor $\tau_0$ in Eq.~(\ref{tau})---the {\em initial} relaxation
events span fewer particles the higher the pressure at which the
spinodal was reached.  This reflects the aforementioned notion that at
a fixed density, higher pressure corresponds with a smaller
coordination number and, hence, relatively open structures. Such open
structures can reconfigure by moving relatively few particles.

One concludes that a colloidal suspension that is quenched starting
from a low density will reach the string-spinodal line from the left
hand side on the diagram in Fig.~\ref{HS} through a series of low
barrier, string-like relaxations. Once the boundary where there can be
a stable aperiodic crystal state is reached, relaxation events
abruptly begin to slow down because they now require activation. The
barrier increase will dramatically extend the lifetime of the
metastable structure compared with the already long timescales needed
for colloidal particles to move about. On times scales shorter than
the so extended time, the system will appear to be arrested and, in
fact, will appear to be jammed since the thermal forces are small
relative to the forces that are imposed externally.

Following an activated reconfiguration event, our argument for the
stress profile suggests the immediate surroundings of the reconfigured
region will experience a partial increase in {\em shear}\/ stress. The
magnitude of the stress then falls off as $1/r^3$ as one moves away
from the rearranging region. Wisitsorasak and
Wolynes~\cite{Wisitsorasak02102012} have shown that externally imposed
shear on glasses leads to an instability. This argument gives
quantitative values for the yield strength of glasses. Thus we see
that if a local aging event occurs in a system that is on the high
density side of the crossover line but not too far away from it, the
surrounding regions will in turn become unstable to forming additional
fractal excitations thus facilitating a series of avalanches
nearby. After reconfiguration, the newly equilibrated regions, now
denser than they were before, will lead to force percolation.
Detailed analysis of these facilitation effects will be a subject of
future work.


\section{The relation to local buckling instability}

We cannot emphasize enough that the relaxation phenomena we have
discussed in this paper are not accessible to strictly meanfield
analyses for two distinct reasons: On the one hand, relaxation events
are local with finite barriers. On the other hand, the relaxations are
truly irreversible and result in either a spatially compact, {\em
  activated} reconfiguration from a metastable aperiodic free energy
minimum to a stable aperiodic minimum or a nearly barrierless process
upon which the system then again finds itself locally in a stable free
energy minimum. For these reasons, the aging relaxations have no
analog when $D = \infty$, where all particles are treated
equivalently. Nevertheless, this does not necessarily mean that
marginal stability effect similar to those inferred for meanfield
liquids could not be observed in finite dimensions.

Because the local environment in an aperiodic structure does not
typically satisfy any point symmetries, one expects that local cages
in an aperiodic crystal should be less isotropic than they are in
periodic crystals. Suppose for the sake of argument that the cage for
a spherical particle is somewhat elongated in one direction. Notice
this would involve endowing the scalar $\alpha$ with an additional
component that is second rank tensor.~\cite{BL_6Spin, BLelast} As
graphically illustrated in Fig.~\ref{jamstrings}(b), the
particle-particle gap along the short axes of the cage can be made
arbitrarily smaller than the gaps to particles along the longer
axis. This implies that the particle in question would be considered
as making contact with only a fraction of its original neighbors, if
jammed to high pressure. If the resulting number of contacts is less
than that required for mechanical stability, the particle's being in
the middle of the cage corresponds to an unstable, hypostatic
configuration, not a mechanically stable local free energy. It follows
that the latter unstable configuration will itself be a transition
state separating two {\em stable} configurations, one of which will be
inevitably chosen thus implying a symmetry breaking transition. This
transition will be accompanied by the formation of a region with
modestly less coordination on the other end of cage. This cage
``buckling'' can be thought of as the initiation of a stringy or
fractal reconfiguration event.

A qualitative argument can be made, see the Supplementary Information,
for the amount of local anisotropy to be expected in the local
cage. This anisotropy depends on the shear modulus of the glass in the
beginning of the quench. The typical spacing between particles can be
expressed through its bulk modulus, as in Eq.~(\ref{SW}). As we
describe in the Supplementary Information, by comparing longitudinal
and transverse displacements one can write down the following,
qualitative criterion for when cages made of the nearest neighbors
would begin to buckle:
\begin{equation} \label{HSbucklingM} \wtK \simeq (9 \pi/2)
  \widetilde{\mu}_f.
\end{equation}
The solution for this equation allows us to plot a one-particle
``buckling instability'' line in Fig.~\ref{HS}. We notice this line
crosses the ``Kauzmann line'' $s_c = 0$ at a finite pressure.

Likewise, larger regions could also undergo {\em higher}-order
buckling transitions when we think of a region having to choose
between two alternative states. Such transitions would however occur
at higher pressures than those causing one-particle buckling
instabilities:
\begin{equation} \label{HSbuckling2M} \wtK \gtrsim \widetilde{\mu}_f
  (L/a)^3,
\end{equation}
where $L$ is the size of the region undergoing the symmetry breaking.

Buckling transitions---which we can see are related to the
reconfiguration events discussed above---clearly result in a splitting
of the original free energy basin into sub-basins. They seem to be the
analog of the higher-order RSB seen in the $D=\infty$ models. This
suggests that the buckling is a finite-dimensional analog of the
Gardner transition.

\section{Fluctuations and dynamical heterogeneity}

We have analyzed the activated events and instabilities in the
amorphous solids in an approximation that accounts for the exponential
multiplicity of glassy aperiodic crystals but that neglects the
fluctuations in specific volume, energy, and configurational entropy
that inevitably arise from this diversity. Xia and
Wolynes~\cite{XWbeta} showed how the fluctuations in driving force for
activated transitions can account for dynamical heterogeneity.  Their
analysis predicted, in quantitative agreement with experiment, the
correlation between the stretching exponent for the alpha relaxation
in equilibrated glassy liquids and liquid fragility. Stevenson and
Wolynes later showed how these driving force fluctuations also act to
smear out the instabilities due to fractal excitations. This smearing
then gives rise to a tail to the barrier distribution referred to as
the secondary or beta relaxations. Their theory also predicts that
beta relaxation events have merged with the alpha relaxation near the
crossover point.

In the moderate pressure regime for amorphous solids these earlier
analyses should suffice but since at high external pressure the
driving force also has a contribution from the volume mismatch energy,
the fluctuations of the latter need to be accounted for at high
pressures. These fluctuations should depend on fluctuations in
specific volume and shear modulus. We leave further analysis of the
fluctuation effects, owing to its complexity, to future work. We note
that fluctuations in the driving force can, if large enough,
completely destroy the random first order transition itself and turn
it into a continuous one. This was discussed by Stevenson et
al.~\cite{stevenson:194505} in their explicit mapping of glassy liquid
landscapes onto magnets with both random fields and random
couplings. Near equilibrium for molecular liquids, their analysis
predicted that the one step RSB transition remains intact. The
fluctuations in driving force, in principle, however could become
large enough to destroy the 1-RSB transition at large pressures. To
get a sense of this effect, we note that by Eqs.~(\ref{dg}) and
(\ref{Kp}) fluctuations in the driving force at the cooperativity size
$N^*$ can be roughly estimated as $\delta \Delta g \simeq p/\sqrt{3
  N^*}$, where we have used the standard expression~\cite{LLstat} for
pressure fluctuations: $(\delta p)^2 = k_B T K_S/V$, where $K_S$ is
the adiabatic bulk modulus and $V$ volume. Clearly, even for the
smallest allowable values of $N^*$ of order one, $\delta \Delta g <
|\delta \Delta g|$. This implies that fluctuations in the driving
force should modify the energetics of activation only quantitatively,
not qualitatively.  In any event, the present argument leaves open the
possibility of an additional region of marginal stability at high
pressures near the densest random-close packing, see also our earlier
remarks regarding the buckling-instability line.

\section{Summary and discussion}

To zeroth order, amorphous solids can be thought of as being quenched
liquids. But when we look more deeply, we see that their properties do
depend on the way they are prepared. In this regard it is important to
realize that in terms of our diagram plotted in terms of dimensionless
quantities, the quenching protocols are typically quite different for
making molecular glasses than they are for preparing amorphous solids
made of larger particles such as colloids or grains. Ultimately these
differences trace back to the enormous size differences between atoms,
colloidal particles, and grains. Molecular glasses, owing to their
being made of rapidly moving atomic-scale constituents, become
sensibly rigid and fall out of equilibrium in the laboratory only deep
in the activated regime, when the activation barriers are quite high,
leading to times fourteen orders of magnitude beyond the basic
molecular scale. In contrast, the intrinsic thermal motions of
colloids or granular objects are already quite slow even when
unactivated because the particles are so large. This means the
emergence of even a very modest barrier is enough to lead to apparent
rigidity on human time scales. Another contrast between molecular
glasses and colloidal assemblies is that the hard-sphere, or kinetic,
part of the pressure is already very large for molecular liquids
before they are quenched but in the total pressure, this hard sphere
contribution is nearly canceled by a very large attractive
contribution due to inter-particle cohesive
forces.~\cite{LonguetHigginsWidom} Thus very high dimensionless
kinetic pressures are very difficult to achieve for molecular liquids
by compression in the laboratory. They typically require at least
megapascals to have any significant effects; such pressures require
specialized apparatus. For colloids or granular assemblies, in
contrast, there is little problem achieving, in the laboratory, very
high dimensionless kinetic pressures because the absolute number
density of the constituents of such assemblies is orders of magnitude
lower in absolute terms, so the compression energy is easy to
maintain. These differences between molecular and colloidal systems
mean that the compression effects highlighted in this paper hardly
show up in laboratory studies or molecular glasses but become
paramount for experiments on granular assemblies or colloids. The high
compression possible for the colloids or granular assemblies means
that "jamming" occurs quite readily for them but does not occur to any
extent for molecular glasses. At the same time compression still can
only barely take a typical colloidal system into the activated
region. This implies amorphous solids prepared from grains or colloids
inevitably end up being near the line of marginal stability but
molecular glasses prepared in the laboratory will generally not be
marginally stable but will be deep in the activated regime.

The string-induced spinodal apparently could \`a priori lead to a
divergent correlation length and, hence, criticality. Any incipient
criticality however would be avoided because of the local activated
events. Indeed, already the dimensionless activation barrier
$F^\ddagger/k_B T$ decreases with pressure for a sample quenched
starting in the landscape regime, as does of course the relaxation
time, which scales exponentially with $F^\ddagger/k_B T$. (There is
also a decrease in the prefactor because the collisional time
decreases with pressure as $1/p$.)  We see that activated events
should destroy long range correlations in the non-equilibrium solid
entirely just as the incipient criticality due to mode-mode coupling
effects is destroyed by the activated reconfigurations in equilibrated
liquids.~\cite{MCT1, LW_soft, BBW2008}

That the incipient long-range correlations are short-circuited by
activated reconfiguration events is consistent with observations on
colloidal suspensions,~\cite{Weitz2006} where it has been possible to
identify large, spatially correlated clusters that percolate the
sample but only on short timescales.  On times beyond a certain
threshold, the clusters do break up. The threshold time can be made
arbitrarily longer than the observation time by sufficiently
compressing the suspension, in agreement with the results of the
present analysis for the emergence of activated barriers upon crossing
the string spinodal. Clearly, a rigidity-percolated colloidal
suspension can be regarded as jammed because its temperature is
effectively zero.

We note that our view of jamming phenomena as a spinodal is consistent
with ``unjamming'' simulations of soft repulsive spheres due to Ikeda
et al.~\cite{ikeda:12A507}, who found a scaling relation between the
correlation length $\xi_j$, relaxation time $\tau_j$, and proximity
$|\varphi_j - \varphi|$ to an unjamming transition. At sufficiently
low temperatures, they find $\xi_j^2 \propto \tau_j \propto |\varphi_j
- \varphi|^{-1/2}$.  Now, in general one finds near a meanfield
spinodal, $\xi_j \sim |\varphi_j -
\varphi|^{-1/4}$.~\cite{ISI:A1978GF11500008} Kirkpatrick and Wolynes
identified this spinodal scaling in their early analysis of glasses in
the Kac limit.~\cite{MCT1} Also assuming no dynamical renormalization,
the Onsager-Landau ansatz for the evolution of an order parameter, say
$\eta$: $\dot{\eta} \propto \prtl F/\prtl \eta$,~\cite{Goldenfeld}
leads to $\tau_j$ being proportional to $\xi^2_j$.  ($F$ is the
appropriate free energy.) This is just what Ikeda et
al.~\cite{ikeda:12A507} find.

Colloidal suspensions differ in an essential way from molecular
liquids in that their dimensionless kinetic pressure is much lower
than the dimensionless pressures that is ordinarily applied by the
experimenter. (This fact is often referred to as such a suspension
being nearly zero temperature, a notion that is even more pertinent to
granular matter.)  Accordingly, the collisional timescale, which is
now set by the viscosity of the solvent and the very large particle
size, is so long that configurational equilibration is intrinsically
much slower than in molecular-scale glassformers, easily by 10 orders
of magnitude.~\cite{L_AP} Thus we conclude that colloidal suspensions
will typically approach the string crossover from the {\em low}
density side, where they become solidified.  Upon compression, the
suspension begins to relax toward high density states at an (osmotic)
pressure greatly exceeding its equilibrium value (but still much lower
than experimenter-inflicted forces!).  Until the string-crossover line
is reached, relaxations are essentially barrierless, being limited
primarily by the viscous drag of the solvent on the individual
colloidal particles. During such relaxation, individual particles will
move distances comparable to the particle spacing. However, once the
crossover is reached, there will now arise activation
barriers,~\cite{LW_soft} which will further slow down the relaxation,
while the particle displacements will become significantly smaller and
comparable to the typical vibrational amplitude. The latter amplitude
decreases inversely proportionally with the pressure, as already
mentioned. Because of the emergence of activation, the suspension will
be largely arrested on experimentally relevant scales beyond but very
near to the spinodal line.

The above discussion of colloidal suspensions is also directly
relevant to computer simulations of hard spheres that are monodisperse
or nearly so. In this case, the ``experimenter's'' time scale must
again be very short out of necessity because monodisperse hard spheres
actually crystallize readily. On the other hand, heterodisperse
mixtures do not crystallize so easily, which allows simulations to go
somewhat deeper in the activated regime. One caveat is that it is not
clear at present to what extent the slow processes in such
polydisperse mixtures could entail phase separation and de-mixing. The
latter effects seem to be particularly important in continuously
heterodisperse mixtures, whose mixing entropy is strictly infinite,
see however Ref.~\onlinecite{doi:10.1063/1.4972525}. We leave this
aspect of the problem for future work.

We have mentioned a way of looking at the symmetry breaking taking
place in quenched glasses as a ``buckling transition.'' Conceivably
from this viewpoint, the low-density, replica-symmetric and
high-density, 1-stage RSB region will be separated by a distinct set
of instabilities that formally correspond with higher-than-one state
RSB. Despite the poetic similarity of the buckling instability with
the meanfield Gardner transition, the connection between the two, if
any, is not yet clear. In any event, it is quite reasonable that at
the spinodal between the replica-symmetric and 1-step RSB sectors,
there is a potential for a continuous replica-symmetry breaking. The
latter notion is consistent with the findings by Dzero et
al.~\cite{DSW2005, PhysRevB.80.024204} that interfaces between
distinct aperiodic free energy minima exhibit higher-order RSB.
Indeed, similar interfaces are expected to separate replica symmetric
and 1-stage RSB regions near the equilibrium crossover.~\cite{XW} Such
interfaces exhibit significant variations in the local value of the
order parameter $\alpha$,~\cite{LW_soft} which are {\em also}
requisite for the buckling instabilities. As a corollary of the above
considerations---and consistent with meanfield results---we conclude
that the spinodal line can also be thought of as a line of
jamming-unjamming transitions. It is possible that the argument for
the buckling instability line could be extended all the way to the
equilibrium crossover, while its separation from the spinodal becomes
progressively smaller on approach to $\varphi_\scr$. Just as for the
stringy spinodal, buckling is not expected to cause true criticality
owing to the activated reconfiguration events.

As pointed out in the beginning of the Section, ordinary glasses made
of organic molecules or inorganic network materials---as opposed to
collections of macromolecular or colloidal assemblies---are generally
quenched starting well in the landscape regime. At the same time, we
have seen that in order to observe the re-emergence of the string
spinodal at high pressure, following a conventional quench, one must
increase the {\em kinetic pressure} by several fold. Elementary
estimates~\cite{L_AP} show that the kinetic pressure in actual
substances, $p \simeq k_B T/a^2 d$, exceeds atmospheric pressure by
about four orders of magnitude. This corresponds to gigapascals. Such
pressures have been achieved in studies of organic
glassformers~\cite{PhysRevLett.92.245702} and we hope that further
experiments along these lines will be performed. According to our
analysis, however, to see the compression-induced instability requires
this pressurization to be sufficiently fast to avoid aging toward more
stable structures.  Once such aging occurs, still higher pressures are
required to observe the spinodal coming from the high density, glassy
side.  In any event, the spinodal obviously cannot be reached using
conventional thermal quenches because in the latter, the ambient
pressure remains comparable to atmospheric pressure. This difficulty
of observing marginal stability comes exclusively from the attractive
interactions in actual molecular systems positioning them already
quite high in the dimensionless kinetic pressure.



Although our arguments were explicitly illustrated in this paper using
hard monodisperse spheres, arguments like those presented here are
also applicable to {\em soft} particles, such as actual molecules or
the soft repulsive spheres mentioned earlier in the context of studies
of unjamming transitions by Ikeda et al.~\cite{ikeda:12A507} Equations
of state for molecular substances can be mapped onto those for hard
spheres; see the Supplementary Information for a qualitative
discussion of correlation between the slope of configurationally
adiabatic quenches and the fragility of the liquid.  We have already
mentioned that the slope of the string spinodal near the equilibrium
crossover is expected to be greater for molecular liquids than for
strictly hard objects. This implies that reaching the spinodal
instability from the landscape side in such liquids is even harder for
systems with soft repulsion than for hard spheres, in light of the
aforementioned difficulties of rapidly creating and then sustaining,
in the laboratory, high pressures in molecular systems. When dealing
with molecules, quantum effects could play a role; these effects
depend on the absolute particle masses, of course. Such quantum
effects are not expected to affect significantly the location of the
string spinodal. Indeed, no quantities entering the condition
(\ref{cond1}) explicitly refer to the particle mass except the
vibrational free energy difference, which would have both an entropic
part and a zero energy contribution. The vibrational free energy
change during isothermal aging---which was dropped from
Eq.~(\ref{cond1}) in the first place---will be negligible. We note
that marginal stability due to buckling will be {\em also} efficiently
suppressed by the quantum effects, see the Supplementary
Information. Finally, at the high pressures requisite for marginal
stability various structural transitions of electronic origin
generally begin to occur.~\cite{ANIE:ANIE200602485} These are beyond
the scope of the present paper.

In view of the virtual inaccessibility of the pressure-induced
spinodal in actual molecular systems and the suppression of the
buckling owing to quantum effects, we conclude that marginal stability
like that manifest in the Gardner transition is not the cause of the
anomalies seen in cryogenic glasses, viz., the two-level systems and
their higher-temperature companion, the Boson peak. Their
universality,~\cite{FreemanAnderson} emphasized early on by Yu and
Leggett,~\cite{YuLeggett} instead has its origin in the low barrier
tail of both compact and stringy reconfiguration events~\cite{LW,
  LW_BP, LW_RMP} characteristic of glassy systems in three dimensions.

It would be interesting to reach the spinodal from the high density
side. One way to do this would be to study a colloidal suspension of
nanoclusters of appropriate size or, perhaps, artificial polymers or
biomolecules so that the intrinsic time scale of motion is neither too
short nor too long. Unfortunately, nanoclusters in this size range
have rather complicated particle-particle interactions.  The
unavoidably large dispersion forces must be compensated by Coulomb
repulsion in order to avoid aggregation, which itself leads already to
rather complicated phase behaviors,~\cite{PWLmeso} such as the
formation of gels.~\cite{Muschol1997} Yet certain proteins, such as
hemoglobin, conceivably may fit the bill.

{\em Acknowledgments}: We gratefully acknowledge early discussions
with Dr. Pyotr Rabochiy. V.L.'s work is supported by the NSF Grants
CHE-1465125 and MCB-1518204 and the Welch Foundation Grant
E-1765. P.G.W.'s work is supported by the Center for Theoretical
Biological Physics sponsored by the National Science Foundation (NSF
Grants PHY-1427654).  Additional support to P.G.W. was provided by the
D. R. Bullard-Welch Chair at Rice University (Grant C-0016).

\bibliography{lowT}

\clearpage

\setcounter{section}{0}
\setcounter{equation}{0}
\setcounter{figure}{0}
\setcounter{table}{0}
\setcounter{page}{1}
\makeatletter
\renewcommand{\theequation}{S\arabic{equation}}
\renewcommand{\thefigure}{S\arabic{figure}}
\renewcommand{\bibnumfmt}[1]{[S#1]}

\begin{widetext}

  \begin{center} {\bf \large {\em Supporting Information}: Locality
      and marginal stability in structural glass: \\ The relations
      among aging, jamming, and the glass transition} \medskip

   {\large Vassiliy Lubchenko$^{1}$ and Peter G. Wolynes$^2$} \medskip

   {\normalsize $^1$Departments of Chemistry and Physics, University
     of Houston, Houston, TX 77204

    $^2$Departments of Chemistry, Physics and Astronomy, and Center
    for Theoretical Biological Physics, Rice University, Houston, TX
    77005}

  \end{center}
  
\end{widetext}

\section{Self-consistent determination of the final state resulting
  from aging of a quenched liquid}

Numerical estimates for all of the features on the phase diagram of a
quenched liquid were obtained for monodisperse hard spheres as the
model system.  For the $s_c > 0 $ part of the equilibrium EOS, we use
the Percus Yevick equation of state:
\begin{equation} \wtp_\text{PY}(\varphi) \equiv \frac{p}{\rho k_B T} =
  \frac{1+\varphi+\varphi^2}{(1 - \varphi)^3}.
\end{equation}
For configurational adiabats, we use the Salsburg-Wood  form
\begin{equation} \label{SWs} \wtp_\text{SW}(\varphi, \varphi_\smax) =
  \frac{3}{\varphi_\smax/\varphi-1} + 1,
\end{equation}
given as Eq.~(\ref{SW}) in the main text. The $s_c = 0$ part of the
equilibrium EOS is given by Eq.~(\ref{SWs}) with $\varphi_\smax =
\varphi_K^\infty 0.64$. Configurationally adiabatic quenches are
assumed to be SW isotherms, whose $\varphi_\smax \le
\varphi_K^\infty$, of course.  According to Rabochiy and
Lubchenko,~\cite{RL_LJ} (RL) the filling fraction at the crossover is
$\varphi = 0.47$, which may well be an underestimate: The filling
fraction at the melting point, according to
simulation,~\cite{HooverRee1968} is $0.49$. The reason for this likely
underestimate is RL's using the Bennett lattice~\cite{Bennett} in
their DFT calculation. This does not pose significant issues, however,
as long as we use the very same DFT setup for the rest of the
quantities. Just such data are available thanks to
L\"owen,~\cite{Lowen} who reports the elastic constants as functions
of the filling fraction for an aperiodic crystal state. For instance,
at the crossover, he obtains $\wtK = 29.8$ and $\widetilde{\mu} =
24.9$. Thus we adopt $\mu/K = 24.9/29.8$ independent of density and
pressure.

With the above in mind, Eqs.~(\ref{Deltaphi}), (\ref{EOS}) and
(\ref{muK}) yield the following system of equations that needs to be
solved to determine self-consistently the density $\varphi_\seq$ and
pressure $\wtp_\seq$ of the aged region, if the initial, quenched
state was at density $\varphi$ and pressure $p$:
\begin{align} \label{dppS} \varphi \wtp - \varphi_\seq \wtp_\seq &=
  (\varphi_\seq - \varphi) \: \wtK_\seq \:
  \frac{2(1-2\sigma)}{1+\sigma} \\
  \wtp &= \wtp_\text{SW}(\varphi, \varphi_\smax) \\
  \wtp_\seq &= \left\{ \begin{array}{ll}
      \wtp_\text{PY}(\varphi_\seq), & \varphi_\seq \le \varphi_K \\
      \wtp_\text{SW}(\varphi_\seq, \varphi_K^\infty), & \varphi_\seq >
      \varphi_K
    \end{array} \right.
\end{align}
Here, the density of the Kauzmann state $\varphi_K$ is defined as the
intersection of the $s_c > 0$ and $s_c = 0$ parts of the equilibrium
EOS. The reduced bulk modulus of the aged states, $\wtK$, is computed
by differentiating the configurational adiabat $p(\varphi)$ at the
target state and depends only on the density of the latter:
\begin{equation} \wtK_\seq \equiv \left. \frac{\prtl}{\prtl \varphi} [
    \varphi \: \wtp_\text{SW} (\varphi, \varphi_\smax^*)) ]
  \right|_{\varphi=\varphi_\seq},
\end{equation}
where the maximum attainable density $\varphi^*$ along the pertinent
SW isotherm is determined by equating the latter with the PY isotherm:
\begin{equation} \wtp_\text{SW} (\varphi_\seq, \varphi_\smax^*) =
  \wtp_\text{PY} (\varphi_\seq).
\end{equation}

As a practical matter, it is much more convenient to determine
$\varphi$ as a function of $\varphi_\seq$ than the other way around
because (a): the SW isotherm affords one to explicitly write out
$\varphi$ and $\varphi_\smax$ as functions of $\wtp$ and (b): the
equilibrium EOS is only piece-wise smooth and experiences a slope
discontinuity. The so obtained families of $\varphi$ corresponding
with five particular values of $\varphi_\seq$ are shown in
Fig.~\ref{HS} of the main text using thin lines with arrows.

To obtain Eq.~(\ref{dpphigh}), which we replicate here:
\begin{equation} \label{dpphighS} \varphi_\smax^* \wtp -
  \varphi_K^\infty \wtp_\seq = \wtp_\seq^2 \: A (\varphi_K^\infty -
  \varphi_\smax^*) \: \frac{2(1-2\sigma)}{1+\sigma},
\end{equation}
we use Eq.~(\ref{dppS}) and note that in the $\wtp, \wtp_\seq \to
\infty$ limit, one has $\varphi \to \varphi_\smax^*$, $\varphi_\seq
\to \varphi_K^\infty$, and the asymptotic expression in Eq.~(\ref{Kp})
applies. Clearly, the term $\varphi_K^\infty \wtp_\seq$ in the
l.h.s. can be neglected in this limit because $\wtp_\seq \propto
\wtp^{1/2}$, as pointed out in the main text. Consequently, one gets
$\Delta g/k_B T \to -\wtp$, and so Eqs.~(\ref{Falpha}) and
(\ref{dppS}) yield Eq.~(\ref{FT}) of the main text. Note that the
asterisk at $\varphi_\smax^*$ is dropped there, without risk of
confusion, to reduce cluttering.

\section{Determination of the slope of the spinodal line near the
  equilibrium crossover}

It is instructive to ask whether the slope of the string spinodal is
greater or less that the slope of the configurational adiabat, near
the equilibrium crossover $(\varphi_\scr, p_\scr^{-1})$. More
specifically, we wish to determine whether or not one will encounter
the string spinodal for conventional quenches starting below the
equilibrium crossover, which approximately follow configurational
adiabats. To settle this, we need to calculate the (logarithmic) slope
of the crossover line,
\begin{equation} K_s \equiv (\prtl p/\prtl \ln
  \varphi)_{p=p_s(\varphi)},
\end{equation}
near the equilibrium crossover $p_{\scr, \seq}$, where $p_s(\varphi)$
is the pressure vs. density dependence specifying the string-spinodal
line.

\begin{figure}[h]
  \includegraphics[width= .7 \figurewidth]{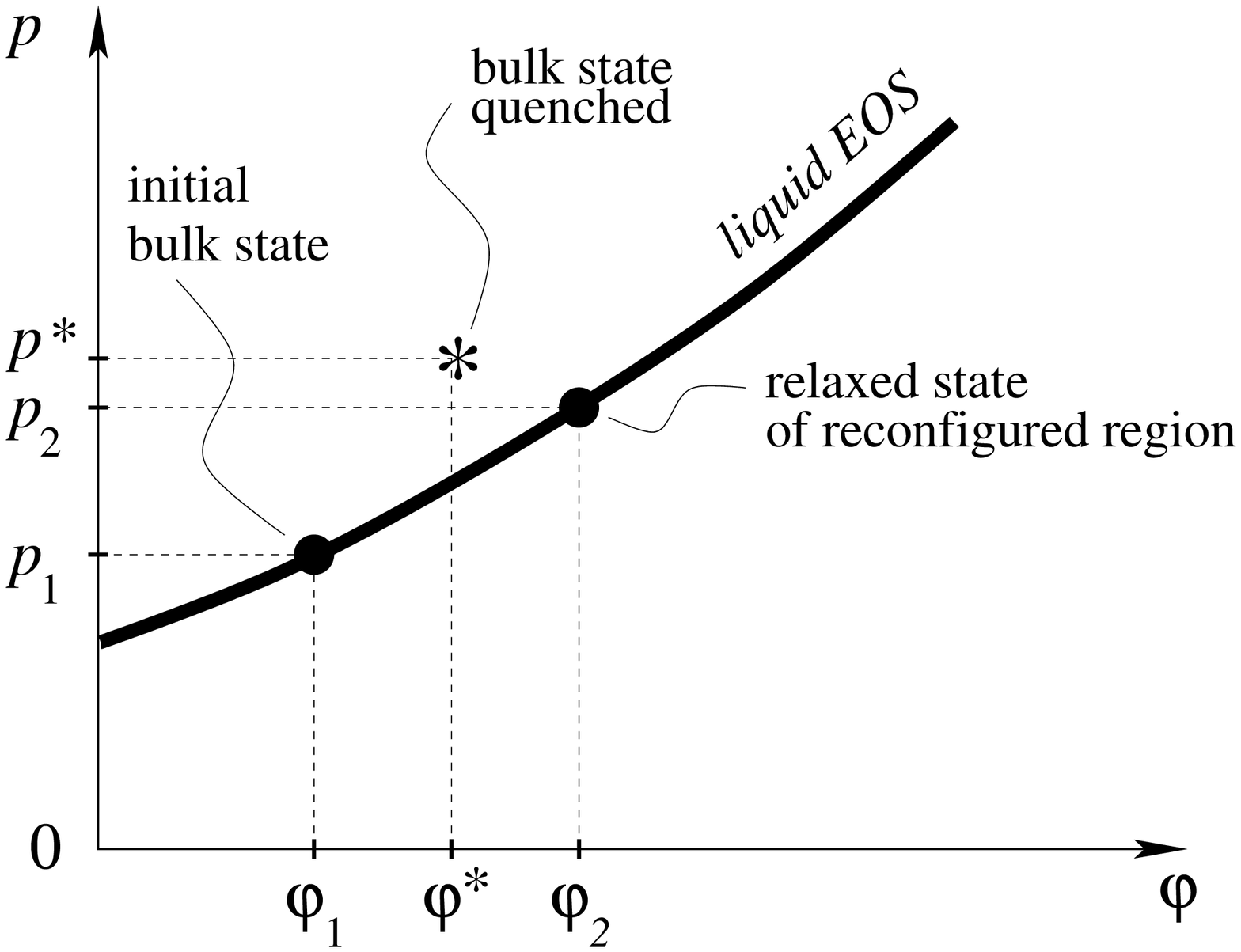}
  \caption{\label{protCross} Illustration of the protocol used to
    determine the ultimate pressure of a reconfigured state starting
    from the a $(\varphi_1, p_1) \to (\varphi^*, p^*)$ quench. Note
    protocols $\varphi^* < \varphi_1$ can be imagined but would be
    difficult to realize in the laboratory.}
\end{figure}

Consider an infinitesimally shallow quenching protocol depicted in
Fig.~\ref{protCross} so that the changes in the filling fraction and
pressure are all infinitesimally small. Denote the ``direction'' of
the quench with $K_q$ so that
\begin{equation} \label{Kq} K_q = \varphi \: \frac{p - p_1}{\varphi -
    \varphi_1},
\end{equation}
where $p_1$ is the starting value of pressure.  The value $p_\seq$ of
the pressure following relaxation from the quenched state at pressure
$p$ is determined by the condition from the main text's
Eq.~(\ref{Deltaphi}):
\begin{equation} \label{Deltaphi2cross} p - p_\seq =
  \frac{\varphi_\seq - \varphi}{\varphi} \: \frac{4\mu}{3}.
\end{equation}
Both point 1 and the equilibrated states belong to the equilibrium
EOS,
\begin{equation} \label{KT} K = \varphi \: \frac{p_\seq -
    p_1}{\varphi_\seq - \varphi_1},
\end{equation}
where $K$ is the equilibrium bulk modulus. Using Eqs.~(\ref{Kq}) and
(\ref{KT}), one can rewrite Eq.~(\ref{Deltaphi2cross}) as
\begin{equation} \label{Deltaphi2cross2} (\varphi - \varphi_1) \wtK_q
  - (\varphi_\seq - \varphi_1) \wtK = \frac{\varphi_\seq -
    \varphi}{\varphi} \:\frac{4\widetilde{\mu}}{3},
\end{equation}
where we have used that all $\Delta \varphi$'s and $\Delta
p$'s are infinitesimally small.

To determine the slope of the crossover line near the equilibrium EOS,
we place state 1 exactly at the equilibrium crossover and set
$p^\text{(s)} = p$ in Eq.~(\ref{cond1}) of the main text:
\begin{equation} p/\rho - p_\seq/\rho_\seq = [\wtgm' - s_c/k_B + \ln
  \Omega + \Delta s_\svibr] k_B T.
\end{equation}
Noting that the r.h.s. in the above equation is equal to zero at the
crossover, the latter equation can be rewritten as:
\begin{equation} \label{ppk2} (\varphi - \varphi_1) \wtK_q -
  (\varphi_\seq - \varphi_1) \wtK = (\varphi_\seq - \varphi_1) \wtK \: B,
\end{equation}
up to $\Delta \varphi^2$, where
\begin{equation} \label{BSI} B \equiv \wtK^{-1} \left\{ \wtp +
    \frac{\prtl}{\ln \varphi}[\wtgm' - (s_c/k_B + \ln \Omega + \Delta
    s_\svibr)] \right\}.
\end{equation}
Upon setting $K_s = K_q$ Eqs.~(\ref{Deltaphi2cross2}) and (\ref{ppk2})
yield 
\begin{equation} \label{Kqcross} K_s = \frac{1+B}{1 - \frac{3K}{4\mu}
    B} \: K,
\end{equation}

We now discuss of Eq.~(\ref{Kqcross}) in the context of conventional
glass-formers made of chemical compounds. Since the ratio $3K/4\mu$ is
numerically close to unity for most such systems, the question of the
slope of the crossover line, near the equilibrium EOS, largely reduces
to establishing the value of the dimensionless quantity $B$. To take
advantage of available data for volume dependence of the bulk modulus,
as in Ref.~\onlinecite{0305-4608-10-7-009}, and using
Eqs.~(\ref{gamma2}) and (\ref{cgamma}), we rewrite Eq.~(\ref{BSI}) as
\begin{equation} \label{BSI1} B = \frac{c_\gamma}{2\wtK^{1/2}}
  \frac{\prtl \ln K}{\prtl \ln \varphi} + \frac{1}{\wtK}\left[\wtp -
    \frac{\prtl}{\ln \varphi} (s_c/k_B + \ln \Omega + \Delta
    s_\svibr)\right].
\end{equation}
According to the compilation of data for crystalline alkaline metals
by Soma,~\cite{0305-4608-10-7-009} the dimensionless quantity $\prtl
\ln K/\prtl \ln \varphi$ exceeds unity but not by very much. We adopt
$\prtl \ln K/\prtl \ln \varphi = 3$, based on Fig.~11 of
Ref.~\onlinecite{0305-4608-10-7-009} for concreteness. The value of
the reduced bulk modulus at the crossover can be estimated using
Rabochiy and Lubchenko's result~\cite{RL_Tcr} that at the crossover,
\begin{equation} \widetilde{\mu}_\scr \approx 5.8 \:
  \frac{5-6\sigma}{2(1-\sigma)},
\end{equation}
where $5.8$ is a numerical constant. Note the fraction varies
relatively little for $0 < \sigma < 1/2$, viz., between 2 and 2.5.
Considering that $c_\gamma \simeq 1/2$, one gets for the first term in
Eq.~(\ref{BSI1}) a numerical value of 0.2 or so.

Next, the ratio $\wtp/K$ is bounded from above by its value for hard
spheres, viz., ca. one-tenth but is much much smaller for soft spheres
and can be neglected, according to the discussion in then main text.
The rate of the decrease of the configurational entropy, $\prtl
s_c/\prtl \ln \varphi$ is about $10 k_B$.~\cite{RL_LJ} The latter
value specifically gives excellent agreement for the relation between
the fragility coefficients at constant pressure and
volume,~\cite{RL_LJ} respectively, and will be adopted here for actual
glassformers. A sense for the magnitude $\prtl s_c/\ln \varphi \simeq
10 k_B$ can be obtained by noting that known glassforming liquids
decrease in volume following crystallization. The entropy of fusion is
roughly $1 k_B$ per particle while the volume change is about
10\%. Thus this term contributes one third or so.  We do not know how
to estimate the rate of change of the ``string entropy'' $\ln \Omega$,
however note that $\ln \Omega$ goes roughly as the log number of
nearest neighbors. The latter quantity is expected to change with
density much less than $s_c$; we will assume it does not vary.  The
term $-\wtK^{-1} \frac{\prtl}{\ln \varphi} \Delta s_\svibr$ is bounded
from above by its value for hard spheres, $3/\wtp$, which is
numerically about one third.

One thus obtains that the quantity $B$ is probably less than one but
not by much: $B \approx 0.6 \ldots 0.9$. At the same time, the ratio
$3K/4\mu = (1+\sigma)/2(1-2\sigma)$ varies between 1 and 4 or so in
the generic $\sigma$ range $0.2 \ldots 0.4$,~\cite{PhysRevB.80.132104}
the lower limit more likely than the latter. Thus we obtain that the
ratio $K_s/K_T$ is of order ten, i.e., significantly exceeds one, but
could be significantly higher and even negative. These qualitative
estimates thus suggest that the string spinodal runs well below the
configurational adiabat originating at the crossover and could even
bend toward lower densities.

\section{High pressure asymptotics of the string spinodal}

Eq.~(\ref{dpphigh}) implies that at high pressures and a finite
$\Delta \varphi$, $p_\seq \propto p^{1/2}$, as pointed out in the main
text. This is incompatible with Eq.~(\ref{cond1}), because $\gamma'
\propto \gamma \propto p_\seq$. Thus, the spinodal line must reach the
fully compressed Kauzmann state in the $p \to \infty$ limit: $\varphi
\to \varphi_K^\infty$. (A fortiori, $\varphi_\seq \to
\varphi_K^\infty$.)  Under these circumstances, Eq.~(\ref{dppS})
yields:
\begin{equation} \label{hp1} \wtp - \wtp_\seq = \frac{\varphi_\seq -
    \varphi}{\varphi_K^\infty} \: \wtK_\seq \:
  \frac{2(1-2\sigma)}{1+\sigma},
\end{equation}
while Eqs.~(\ref{cond1}), (\ref{gamma2}) and (\ref{cgamma}) imply
\begin{equation} \label{hp2} \wtp - \wtp_\seq = c_\gamma
  \wtK_\seq^{1/2}.
\end{equation}

Equating the right-hand-sides of Eqs.~(\ref{hp1}) and (\ref{hp2}),
together with the high-pressure limit of the SW isotherm,
\begin{equation} \wtK_\seq = \frac{3
    (\varphi_K^\infty)^2}{(\varphi_K^\infty - \varphi_\seq)^2},
\end{equation}
immediately makes it clear that the density change for an aging event
starting at the spinodal, in the $p \to \infty $ limit, is
proportional to the proximity of the equilibrium state to the
infinitely compressed Kauzmann state: $(\varphi_\seq - \varphi)
\propto (\varphi_K^\infty - \varphi_\seq)$. That is, the $p \to
\infty$ end of the spinodal line is at the infinitely compressed
Kauzmann state. Finally, by employing the SW equation of state, one
readily obtains:
\begin{equation} \label{highPasym} \wtp^{-1} = (\varphi_K^\infty -
  \varphi) \left[\varphi_K^\infty \left(c_\gamma/c_1 3^{1/2}+1
    \right)\left(3+c_\gamma 3^{1/2}\right)\right]^{-1},
\end{equation}
where
\begin{equation} c_1 \equiv \frac{2(1-2\sigma)}{1+\sigma}.
\end{equation}
Clearly the linear dependence specified by Eq.~(\ref{highPasym}) has a
smaller slope than that of the Kauzmann line near $p^{-1} = 0$. The
latter line is given by $\wtp^{-1} = (\varphi_K^\infty -
\varphi)/3\varphi_K^\infty$, according to Eq.~(\ref{SWs}).

\section{The buckling instability}

Here we discuss in some detail another, significantly milder type of
instability that could take place in a compressed aperiodic crystal.
The onset of symmetry breaking and, hence, marginal stability will
take place just when the two sides of the elongated cage in
Fig.~\ref{jamstrings}(b) become distinguishable. Since the coordinate
of a particle can be specified only within its vibrational
displacement, $\alpha^{-1/2}$, one may restate the latter notion more
formally in the form of the following criterion: The symmetry breaking
will occur when the difference in the vibrational displacement between
a long and short axis of the cage exceeds the typical displacement:
\begin{equation} \label{dalpha} \delta (\alpha^{-1/2}) >
  \alpha^{-1/2}.
\end{equation}

Below the crossover, when the cages are relatively stable and
isotropic, the material behaves locally like an elastic solid.  Under
these circumstances, the l.h.s. is essentially the deviation of the
aspect ratio of the cage from unity, times the lattice spacing.  The
latter deviation, in turn, can be thought of as a frozen-in
fluctuation of shear.~\cite{BL_6Spin, BLelast} The magnitude of the
typical strength of local fluctuations of shear can be estimated using
continuum mechanics, by which the shear part of the free energy is
given by~\cite{LLelast} $\int dV \mu (u_{ik} - \delta_{ik} u_{jj})^2$,
where $u_{ik}$ is the strain tensor and $\mu$ the shear modulus. Note
the traceless tensor $(u_{ik} - \delta_{ik} u_{jj})$ exclusively
reflects the shear component of the strain. Hence we adopt $\delta
(\alpha^{-1/2}) \simeq \la (u_{ik} - \delta_{ik} u_{jj})^2 \ra^{1/2}$.
Since shear is responsible for two degrees of freedom, the latter free
energy will typically fluctuate by $k_B T$, yielding:
\begin{equation} \label{dalpha1} \delta (\alpha^{-1/2}) \simeq a (k_B
  T_f/\mu_f V)^{1/2},
\end{equation}
where $\mu_f$ and $T_f$ stand, respectively, for the shear modulus and
temperature at the beginning of the quench.  The label ``$f$'' refers
to ``fictive,'' in deference to the terminology accepted in the field
of glass aging. Note that for compression-induced quenches, $T_f$ is
equal to the ambient temperature $T$, by construction. For a cage
formed by the first coordination shell, the pertinent volume $V$ can
be estimated as $(4\pi/3) (3a/2)^3$, where $a$ is the volumetric size
of a rigid molecular unit that is not significantly perturbed by
liquid motions, or ``bead.''~\cite{LW_soft} At the same time, the
typical vibrational displacement in our solid, i.e. the r.h.s. of
Eq.~(\ref{dalpha}), can be estimated using the equipartition
theorem. For temperature well above the Debye temperature $T_D$ the
displacement is thus simply given by the quantity $(k_B T/K
a^3)^{1/2}$ but should saturate at $(k_B T_D/K a^3)^{1/2}$ with
lowering the temperature.  With these notions in mind, we can rewrite
Eq.~(\ref{dalpha}) as
\begin{equation} \label{dalpha2} \frac{k_B T_f \rho_f}{\mu_f} >
  \frac{9 \pi}{2} \: \max\left(\frac{k_B T \rho}{K}, \frac{k_B T_D
      \rho}{K}\right) \hspace{5mm} \alpha \gg 1,
\end{equation}
where, note, the bulk modulus $K$ and density $\rho$ are
temperature/pressure dependent.  In view of the facts that $9\pi/2
\approx 14$ and that the bulk and shear modulus for most substances
are numerically similar, the above condition can be satisfied only if
one can achieve a very substantial increase in the bulk modulus by
quenching.

That the condition (\ref{dalpha2}) cannot be realized in conventional
glassformers is brought home by the empirical notion that the Debye
temperature in the latter materials is correlated with the melting
temperature~\cite{GrimvallSjodin} and, hence, the glass transition
temperature,~\cite{10.1063/1.2244551, LW_soft} so as not be less than
a quarter of the glass transition temperature, if not more. At the
same time, the elastic constants of solids increase only modestly upon
cooling at constant pressure.~\cite{RWLbarrier, LRactivated} Thus we
conclude that in conventional glassformers made by thermally quenching
from the aperiodic crystal significantly below $T_\scr$, quantum
fluctuations will destroy the type of symmetry breaking associated
with the anisotropy of the immediate coordination shell. In physical
terms, this simply means that because of the softness of inter-atomic
interactions, local idiosyncrasies in the coordination shell are
effectively averaged over as a result of vibrations. {\em Pressure}
induced quenches are much harder to realize in conventional
glassformers and are performed relatively
rarely.~\cite{PhysRevLett.92.245702} Pressures achieved in such
quenches are, numerically, only a fraction of the elastic moduli,
again implying the condition (\ref{dalpha2}) is not met in such
experiments.

A very contrasting situation is presented by {\em rigid} particles,
which are intrinsically classical effectively implying $T_D = 0$.
Here the typical vibrational displacement can be made arbitrarily
small while the bulk modulus can be arbitrarily large. Thus the
buckling instability occurs when:
\begin{equation} \label{HSbuckling} \wtK \simeq (9 \pi/2)
  \widetilde{\mu}_f,
\end{equation}
up to a factor of order one.  The solution of the above equation for
monodisperse hard spheres is shown as the red solid line in
Fig.~\ref{HS}, where we assumed the Poisson ratio of the equilibrated
liquid is density-independent, see details in the Supporting
Information. We observe that the ``buckling instability'' line merges
rather smoothly with the string-spinodal line. Clearly, sufficiently
close to the spinodal the argument leading to Eq.~(\ref{dalpha1}) is
not externally consistent, nor would the string and buckling
instabilities be easily distinguishable in this case.

Eq.~(\ref{dalpha1}) can be applied to regions of size $N$ larger than
the immediate coordination shell of a single particle.  Assuming the
vibrational displacement is distributed evenly across the region, one
obtains a more general statement:
\begin{equation} \label{HSbuckling2} \wtK \gtrsim \widetilde{\mu}_f N,
\end{equation}
c.f. Eq.~(\ref{HSbuckling}), and we are continuing working with rigid
particles. The above equation implies that the ``buckling
instability'' line in Figs.~\ref{nonMF} and \ref{HS} corresponds with
a threshold limit for the quantity $\wtK$ above which such symmetry
breakings that could occur within a continuous spectrum of
lengthscales. At the threshold, the buckling events occur only very
locally, viz., at a single cage size, as just discussed. Above the
threshold, on the other hand, the events could become significantly
more extended. Indeed, in view of Eqs.~(\ref{Kp}) and (\ref{Kred}),
Eq.~(\ref{HSbuckling2}) yields the following scaling relations:
\begin{equation} \label{L} L \propto p^{2/3} \propto T^{-2/3},
\end{equation}
where $L \propto N^{1/3}$ is the physical extent of the ``buckled''
region.  One may say that the diverging length scale above is that
corresponding to jamming a macroscopically stable aperiodic
crystal. To avoid confusion, we note that observable consequences of
the growing correlation length (\ref{L}) will be somewhat limited
since the spatial concentration of the so jammed modes goes as $1/L^3
\propto 1/p^2 \propto T^2$. Interestingly, the corresponding entropy
per unit volume: $L^{-3} \ln 2$---assuming there are two choices for
an individual symmetry breaking event---scales quadratically with
temperature:
\begin{equation} S/V \simeq T^2 \frac{\widetilde{\mu}_f \rho^3 \ln 2}{
    A p^2},
\end{equation}
and likewise for the corresponding heat capacity $C = T(\prtl S/\prtl
T)$.  This is a slower scaling than Debye's $T^3$ law for the phonon
heat capacity. (For rigid objects, the comparison must be made with
the $T$-independent Dulong-Petit law.) In any event, the correlation
length $L$ will stop growing when the strings begin to nucleate, of
course, while removing the above contribution to the total entropy of
the sample. At the same time, of all of the buckling-induced jamming
will be removed.

\section{Slope of a configurationally-adiabatic $p-V$ curve relative
  to its value for the equilibrium isotherm, and its relation to the
  liquid's fragility: A qualitative discussion}

First we focus exclusively on the configurational degrees of freedom.
For any degree of freedom, the adiabatic compressibility is always
less than its isothermal counterpart.~\cite{LLstat} Insofar as one may
regard the configurational degrees of freedom as decoupled from the
vibrations, we may use the standard formula~\cite{LLstat} $(\prtl
V/\prtl p)_S/(\prtl V/\prtl p)_T=C_v/C_p$ to estimate the ratio of the
adiabatic and isothermal bulk moduli in terms of the configurational
parts of the heat capacity:
\begin{equation} \label{pvratio1} \frac{(\prtl p/\prtl
    V)^\text{(conf)}_{S_c}}{(\prtl p/\prtl V)^\text{(conf)}_T} =
  \frac{\Delta C_p}{\Delta C_v},
\end{equation} 
where ``(conf)'' refers to ``configurational'' and it is understood
that $(\prtl p/\prtl V)^\text{(conf)}_{S} \equiv (\prtl p/\prtl
V)^\text{(conf)}_{S_c}$. In writing down Eq.~(\ref{pvratio1}), we
assume that the bulk moduli corresponding to the configurational and
vibrational degrees of freedom are additive, not the
compressibilities:
\begin{equation} \label{pvtotal} (\prtl p/\prtl V) = (\prtl p/\prtl
  V)^\text{(conf)} + (\prtl p/\prtl V)^\text{(vibr)}
\end{equation}
The analogy is with two springs that are connected in parallel, not
consecutively.

The ratio on the r.h.s. in Eq.~(\ref{pvratio1}) can be estimated using
an approximate connection with the fragility index
$m^\text{fr}$:~\cite{XW} $\Delta C \propto m^\text{fr}$ and the
approximate relation between fragility coefficient at constant
pressure and volume:~\cite{RL_LJ, L_AP} $m_p^\text{fr} \simeq
m_V^\text{fr} + 30$. Hence,
\begin{equation} \label{pvratio2} \frac{(\prtl p/\prtl
    V)^\text{(conf)}_{S_c}}{(\prtl p/\prtl V)^\text{(conf)}_T} \simeq
  \frac{m_p^\text{fr}}{m_V^\text{fr}} = \frac{m_V^\text{fr} +
    30}{m_V^\text{fr}} > 1.
\end{equation} 
Finally, Eqs.~(\ref{pvtotal}) and (\ref{pvratio2}) imply that
\begin{align} \frac{(\prtl p/\prtl V)_{S_c}}{(\prtl p/\prtl V)_T} &=
  \frac{(\prtl p/\prtl V)^\text{(conf)}_{S_c} + (\prtl p/\prtl
    V)^\text{(vibr)}_{S_c}}{(\prtl p/\prtl V)^\text{(conf)}_T + (\prtl
    p/\prtl V)^\text{(vibr)}_T} \\ &> \frac{(\prtl p/\prtl
    V)^\text{(conf)}_{S_c} + (\prtl p/\prtl V)^\text{(vibr)}_T}{(\prtl
    p/\prtl V)^\text{(conf)}_T + (\prtl p/\prtl V)^\text{(vibr)}_T} \\
  &> 1 \label{pvineq}
\end{align}

The $m_V^\text{fr}/m_p^\text{fr}$ ratio approaches zero for the
strongest known substances but only marginally smaller than one for
very fragile substances.~\cite{PhysRevE.72.031503} Thus we expect
inequality (\ref{pvineq}) and, hence, the slope disparity between the
EOS and $s_c = \text{const}$ quenches to be greater for stronger
substances.

\end{document}